\documentclass[a4paper,11pt]{article}
\usepackage{graphicx,amssymb,amstext,amsmath}
\usepackage{color, yfonts, tcolorbox}
\usepackage{floatflt}

\usepackage{float}
\usepackage{esint}
\usepackage{subcaption}
\usepackage{mathtools}
\usepackage{commath}
\usepackage{array}
\usepackage{boldline}
\usepackage{multirow}
\usepackage{arydshln}
\usepackage{bigdelim}
\usepackage{hyperref}
\usepackage{simpler-wick}
\usepackage[sorting=none]{biblatex}
\definecolor{cbl}{rgb}{0,0,1}                

\newcommand{\bc}{\begin{center}}
\newcommand{\ec}{\end{center}}
\def\ba#1{\begin{array}{#1}\displaystyle}
\newcommand{\ea}{\end{array}}

\newcommand{\beq}{\begin{equation}}
\newcommand{\eeq}{\end{equation}}
\newcommand{\beqa}{\begin{eqnarray}}
\newcommand{\eeqa}{\end{eqnarray}}

\newcommand{\bi}{\begin{itemize}}
\newcommand{\ei}{\end{itemize}}
\newcommand{\ap}{\textfrak{a}_p}
\newcommand{\OCA}{\color{green}}
\newcommand{\bp}{\textfrak{b}_p}

\newcommand{\bra}{\langle}
\newcommand{\ket}{\rangle}

\newcommand{\N}{{\mathbb{N}}}

\newcommand{\Tr}{{\rm Tr}}

\newcommand{\TT}{{\cal T}}

\def \be {\begin{equation}} 
\def \ee {\end{equation}} 
\def \l {\left(} 
\def \r {\right)} 
\def \la {\langle} 
\def \ra {\rangle}  
\def\lc#1{{ \color{red}  #1}}
\def \blue#1{{ \color{blue}  #1}}
\def \mm#1{{ \color{blue}  #1}}

\begin{document}
\begin{titlepage}
\vspace{0.2cm}
\begin{center}

{\large{\bf{Symmetry Resolved Entanglement of Excited States in Quantum Field Theory III: Bosonic and Fermionic Negativity}}}

\vspace{0.8cm} 
{\large Luca Capizzi{\LARGE $^{\star}$}, Michele Mazzoni$^\spadesuit$, and Olalla A. Castro-Alvaredo$^\heartsuit$}

\vspace{0.8cm}
{\small
{\LARGE $^{\star}$}  SISSA and INFN Sezione di Trieste, via Bonomea 265, 34136 Trieste, Italy\\

\medskip
$^{\spadesuit,\heartsuit}$ Department of Mathematics, City, University of London, 10 Northampton Square EC1V 0HB, UK\\
\medskip

}
\end{center}

\medskip
\medskip
\medskip
\medskip

In two recent works, we studied the symmetry resolved R\'enyi entropies of quasi-particle excited states in quantum field theory. We found that the entropies display many model-independent features which we discussed and analytically characterised. In this paper we extend this line of investigation by providing analytical and numerical evidence that a similar universal behavior arises for the symmetry resolved negativity. In particular, we compute the ratio of charged moments of the partially transposed reduced density matrix as an expectation value of  twist operators. These are ``fused" versions of the more traditionally used branch point twist fields and were introduced in a previous work. The use of twist operators allows us to perform the computation in an arbitrary number of spacial dimensions. We show that, in the large-volume limit, only the commutation relations between the twist operators and local fields matter, and computations reduce to a purely combinatorial problem. We address some specific issues regarding fermionic excitations, whose treatment requires the notion of partial time-reversal transformation, and we discuss the differences and analogies with their bosonic counterpart. We find that although the operation of partial transposition requires a redefinition for fermionic theories, the ratio of the negativity moments between an excited state and the ground state is universal and identical for fermions and bosons as well as for a large variety of very different states, ranging from simple qubit states to the excited states of free quantum field theories.
Our predictions are tested numerically on a 1D Fermi chain. 

\noindent 
\medskip
\medskip
\medskip
\medskip


\vfill
\noindent 
{\LARGE $^{\star}$} lcapizzi@sissa.it\\
$^\spadesuit$ michele.mazzoni.2@city.ac.uk\\
{$^\heartsuit$}o.castro-alvaredo@city.ac.uk

\hfill \today

\end{titlepage}

\section{Introduction}
Over the past two decades, entanglement measures have been widely studied in the context of low-dimensional quantum field theory, starting with several seminal works \cite{HolzheyLW94,Calabrese:2004eu,Jin,Calabrese:2005in,Latorre1,Latorre2,Latorre3} which focused on one measure (the entanglement entropy \cite{bennet}) and on one type of theory, largely 1D conformal field theory (CFT) and its discrete counterpart, critical spin chains. From these papers sprang several important ideas and techniques which have been extensively exploited thereafter. Notable among them are the numerical and analytical observation that the entanglement entropy exhibits universal properties, i.e. properties that depend  only on the theory's universality class characterised by the central charge $c$ and, at the technical level, that conformal symmetry in 1D is itself a powerful computational tool. An important idea to emerge from \cite{Calabrese:2004eu,Calabrese:2005in} and later reinterpreted and generalised to non-critical theories in \cite{entropy} is that entanglement measures can be written in terms of correlation functions of local fields of the quantum field theory (QFT) under study, or, more precisely, of a replica version thereof. 

One particular development of these ideas has been the proposal and study of new measures of entanglement, each tailored to capturing particular features of entanglement and/or of the state whose entanglement is being measured. One such new measure is the (logarithmic) negativity \cite{AEPW,Volume,ep-99,vw-02,Plenio-05,Perratum,PhD} which we now introduce. 
Let us consider a tripartite system consisting of subsystems $A, B$ and $C$. Let us assume that the Hilbert space of system is factorised as
\be
\mathcal{H}_{A\cup B \cup C} = \mathcal{H}_{A}\otimes \mathcal{H}_{B} \otimes \mathcal{H}_{C}\,.
\ee
We may now consider a state in this Hilbert space. This could be a mixed state (such as a thermal state) or a pure state: in this paper, we will focus on the latter. We now ask: what is the entanglement of $A$ with respect to $B$ given the presence of $C$? In other words: what is the bipartite entanglement between two non-complementary regions of a quantum system? The answer will depend on the entanglement measure and on the chosen state. If the state is not factorised, that is, there is entanglement, then the answer will be non-trivial and can be measured by the (logarithmic) negativity.  Let $\rho_{A\cup B}$ be the reduced density matrix (RDM) associated with the subsystem $A\cup B$, resulting from tracing out the degrees of freedom of $C$. If $|\Psi\ket$ is the pure state of the whole system, then
\beq 
\rho_{A\cup B}:=\mathrm{Tr}_C(|\Psi\ket \bra \Psi|)\,.
\eeq 
In order to define the logarithmic negativity we first need to introduce the partially transposed version of $\rho_{A \cup B}$, denoted by $\rho^{T_B}_{A \cup B}$, as follows. We first pick a basis $\{|i\ket_A\},\{|j\ket_B\}$ for $\mathcal{H}_A$ and $\mathcal{H}_B$ respectively. Then, given the expansion of $\rho_{A\cup B}$ in that basis
\be
\rho_{A\cup B} = \sum_{i_A,i^\prime_A,j_B,j^\prime_B}| i_A, j_B\ket \bra i_A, j_B | \rho_{A\cup  B}|i^\prime_A, j^\prime_B \ket  \bra i^\prime_A, j^\prime_B |\,, 
\ee
we require that
\be
\label{eq:bosonic_transp}
\rho^{T_B}_{A\cup B} = \sum_{i_A,i^\prime_A,j_B,j^\prime_B} | i_A, j^\prime_B\ket \bra i_A, j_B | \rho_{A\cup B}|i^\prime_A, j^\prime_B \ket  \bra i^\prime_A, j_B |\,. 
\ee

It is possible to show that $\rho^{T_B}_{A\cup B}$ has real spectrum, but in general it is not positive semi-definite. As a matter of fact, the presence of negative eigenvalues is a signal of quantum entanglement between $A$ and $B$, which can be quantified by the {\it logarithmic negativity}
\be\label{eq:def_negativity}
\mathcal{E} \equiv \log \text{Tr}(|\rho^{T_B}_{A\cup B}|)\,.
\ee
Here the trace is understood to be over the Hilbert space associated with $A\cup B$ and $|\cdot|$ represents the absolute value of an operator (that is $|\mathcal{O}| \equiv \sqrt{\mathcal{O}^\dagger \mathcal{O}}$).
While its direct evaluation may be hard, it has been pointed out that the moments of $\rho^{T_B}_{A\cup B}$, written as $\text{Tr}((\rho^{T_B}_{A\cup B})^n)$ with $n \in \mathbb{N}$,  have a direct field theoretic description \cite{cct-12,cct-13} and they are easier to compute. Then, one can formally recover the logarithmic negativity as
\be
\mathcal{E} = \underset{n_e\rightarrow 1}{\lim} \log \text{Tr}((\rho^{T_B}_{A\cup B})^{n_e})\,,
\label{repneg}
\ee
where the limit is over analytically continued moments for $n_e$ even.

\medskip
An interesting issue that is specific to the logarithmic negativity is the fact that the definitions above apply directly to spin chains or bosonic systems, but are ill-suited to treat fermionic systems. The reason for this is rather technical and can be explained in different ways. Unlike for bosons, the (standard) partial transpose of the Gaussian density matrix of a free fermion state is not Gaussian, which makes the computation of the negativity spectrum particularly difficult. There have been several proposals as to how to modify the definition of $\mathcal{E}$ in a way that is better adapted to deal with fermionic degrees of freedom. 
The first definition of partial transposition, specifically modified for fermionic states, was introduced in \cite{eisler2016entanglement}. However, in (\cite{ssr-17}, Appendix A) it was proved that, because of the anticommuting nature of the fermionic degrees of freedom, two of the standard requirements of a partial transposition operation, namely that if $\rho \equiv \rho_{A\cup B}$:
\beq 
(\rho^{T_A})^{T_B}= \rho^T \quad \mathrm{and} \quad 
\rho_1^{T_A}\otimes \dots \otimes \rho_n^{T_A} = (\rho_1\otimes\dots \otimes \rho_n)^{T_A}\,,
\eeq 
with $T$ representing transposition over the total space (here $T=T_{A\cup B}$),
may not hold with the definition given in \cite{eisler2016entanglement}. On the other hand, these properties are satisfied with the definition introduced in \cite{ssr-17}: this is the \textit{ time-reversal} (or fermionic) negativity, which accounts for the locality properties of fermions \cite{ssr-17,sr-19} and which we present below. Following \cite{sr-19}, we now choose an occupation-number basis for the Hilbert space
\be
|\{n_j\}_A,\{n_j\}_B\ket
\ee
such that all the $n_j\in\{0,1\}$. Then, given the RDM
\be
\rho_{A\cup B} = \sum_{\substack{\{n_j\}_A,\{n_j\}_B \\ \{n'_j\}_A,\{n'_j\}_B}} | \{n_j\}_A,\{n_j\}_B\ket \bra \{n_j\}_A,\{n_j\}_B|\rho_{A\cup B}| \{n'_j\}_A,\{n'_j\}_B\ket  \bra \{n'_j\}_A,\{n'_j\}_B|\,,
\ee
we define the fermionic partial transposition as 
\be\label{eq:ferm_transp}
\rho^{R_B}_{A\cup B} =  \sum_{\substack{\{n_j\}_A,\{n_j\}_B \\ \{n'_j\}_A,\{n'_j\}_B}} i^{\phi(\{n_j\},\{n'_j\})} | \{n_j\}_A,\{n'_j\}_B\ket \bra \{n_j\}_A,\{n_j\}_B|\rho_{A\cup B}| \{n'_j\}_A,\{n'_j\}_B\ket  \bra \{n'_j\}_A,\{n_j\}_B|\,,
\ee
where  $\phi(\{n_j\},\{n'_j\})$ is given by
\be
\phi(\{n_j\},\{n'_j\}) = (\tau_B+\tau'_B)(\text{mod} \,2) + 2(\tau_A+\tau'_A)(\tau_B+\tau'_B)\,,
\ee
and $\tau_{A/B} = \sum_{j\in  A/B}n_j, \tau'_{A/B} = \sum_{j\in  A/B}n'_j$ are the numbers of occupied states in each subsystem. Thus, the novelty of the fermionic partial transposition \eqref{eq:ferm_transp}, as compared to \eqref{eq:bosonic_transp}, is the presence of an additional phase shift which depends on the number of fermions. While in general $\rho^{R_B}_{A\cup B}$ is no longer hermitian for fermionic systems, one can still define the (fermionic) logarithmic negativity as \eqref{eq:def_negativity} by writing the absolute value as
\beq |\rho^{R_B}_{A\cup B}| = \sqrt{(\rho^{R_B}_{A\cup B})^\dagger\rho^{R_B}_{A\cup B}}\,,
\label{Rtrace}
\eeq 
and observing that it is a positive semi-definite matrix.

\medskip
Let us now add the final layer of definitions by introducing symmetry resolved entanglement measures. 
Such measures have become very popular in the past few years and extend the standard definitions by exploiting the presence of internal symmetries. Some of the earliest studies (see \cite{GS,XAS} for the CFT/QFT and quantum spin chain constructions) focused on the entanglement entropies but more recently also the logarithmic negativity has been generalised in a similar fashion \cite{cornfeld2018imbalance,murciano2021symmetry,Chen-22, Gaur-22}. Let us consider a theory with a global $U(1)$ symmetry (i.e. a complex free boson/fermion). In that case, a global $U(1)$ charge $Q_{A\cup B} = Q_A + Q_B$ commutes with the state $\rho_{A \cup B}$
\be
[\rho_{A \cup B},Q_A+Q_B] =0\,.
\ee
Then, it has been shown \cite{cornfeld2018imbalance} that the charge imbalance $Q_A-Q_B$\footnote{To be precise, the correct relation is $[\rho^{T_B}_{A\cup B},Q_A-Q^T_B]=0$, however, the charge imbalance operator is basis independent and in the occupation number basis $Q^T_B = Q_B$ so we can drop the transposition. As earlier, $T$ represents transposition over the full relevant space, in this case $B$.} commutes with $\rho^{T_B}_{A\cup B}$
\be
[\rho^{T_B}_{A\cup B},Q_A-Q_B]=0,
\ee
and it generates a $U(1)$ symmetry for the (bosonic or fermionic) partial transpose. At this point, it is natural to consider the charged moments of the partial transpose
\be
\label{boson_ch_moments_def}
\text{Tr}\l (\rho^{T_B}_{A\cup B})^{n} e^{2\pi i \alpha (Q_A-Q_B)}\r, \quad  \alpha \in [-1/2,1/2]\,,
\ee
as measures of the {\it symmetry resolved entanglement negativity}, generalising the standard moments ($\alpha=0$). For fermions, the definition above is changed to
\be
\label{fermion_ch_moments_def}
\text{Tr}\l |\rho^{R_B}_{A\cup B}|^{n} e^{2\pi i \alpha (Q_A-Q_B)}\r, \quad  \alpha \in [-1/2,1/2]\,.
\ee
The computation of charged moments of the partial transpose was performed in \cite{murciano2021symmetry} in the ground and thermal state of massless free fermions in 1+1 dimensions, where the universal UV divergences were captured by the underlying CFT. They have also been measured in an experimental set up in  Ref. \cite{neven-2021}. In this work, we are interested in zero-density quasi-particle states of (massive) QFT, obtained as excitations of the ground state with finite number of particles at given momenta. We aim to compute the contribution to the charged moments given by the quasi-particles, which arises in addition to the zero-point fluctuations. These states are ``zero-density" in the sense that they contain a fixed and finite number of quasi-particle excitations within an infinite volume. The present work applies to the same kind of excited states considered in \cite{PRLexcited,castro2018entanglement,castro2019entanglement,graphpaper} for the standard entanglement measures and more recently in \cite{partI,partII} for the symmetry resolved entropies. 

\medskip

We will now briefly state the main results of this work. Let us consider a QFT in $d+1$ dimensions carrying a global $U(1)$ symmetry and two non-complementary spacial regions $A$ and $B$. We take the vacuum state $|0 \ket$, an excited state $|\boldsymbol{k}\ket $ containing $k$ identical quasi-particle excitations with unit charge, and we construct the associated reduced density matrices (RDM) over $A \cup B$ as
\be
\rho_{A\cup B,0} \equiv \text{Tr}_{C}(|0 \ket \bra 0 |), \quad \rho_{A\cup B} \equiv \text{Tr}_{C}(|\boldsymbol{k} \ket \bra \boldsymbol{k}|)\quad \mathrm{with}\quad C\equiv {\overline{A\cup B}}\,,
\ee
for the ground state $|0\ket$ and excited state $|\boldsymbol{k}\ket$, respectively.
We consider the limit when the generalised volume $V_A, V_B, V_C$ of each region goes to infinity while the ratios
\beq 
r_A:=\frac{V_A}{V}, \quad r_B:=\frac{V_B}{V} \quad \mathrm{and} \quad r:=\frac{V_C}{V}=1-r_A-r_B\,,
\eeq
are finite and $V:=V_A+V_B+V_C$. We then define the ratio of charged  moments:
\beq 
\mathcal{R}^n_k(r_A, r_B, r; \alpha):=\frac{\text{Tr}\l (\rho^{T_B}_{A\cup B})^n e^{2\pi i \alpha (Q_A-Q_B)}\r}{\text{Tr}\l (\rho^{T_B}_{A\cup B,0})^n e^{2\pi i\alpha (Q_A-Q_B)}\r}\,.
\eeq 
We find this ratio to be universal (UV finite), and for a single particle excitation ($k=1$) given by \beqa 
\label{eq:ratio_ch_mom_abs}
\mathcal{R}_1^n(r_A, r_B, r; \alpha) = e^{2\pi i \alpha}r^n_A + e^{-2\pi i \alpha}r^n_B  +\l \frac{r+\sqrt{r^2+4r_Ar_B}}{2}\r^{n} +\l \frac{r- \sqrt{r^2+4r_Ar_B}}{2}\r^{n}\,.
\eeqa
We notice that the last in \eqref{eq:ratio_ch_mom_abs} is positive/negative when $n$ is an even/odd integer and therefore two distinct analytic continuations over the even/odd integers are present. In particular, the analytic continuation from $n$ even to $n=1$ gives 
\beqa 
\label{analy}
\lim_{n\rightarrow \frac{1}{2}}\mathcal{R}_1^{2n}(r_A, r_B, r; \alpha) = e^{2\pi i \alpha}r_A + e^{-2\pi i \alpha}r_B  + \sqrt{r^2+4r_Ar_B}\,.
\eeqa
Comparing to the $\alpha=0$ result found in \cite{castro2019entanglement,graphpaper} we see that the phases $e^{\pm 2\pi i \alpha}$ only enter some of the factors in (\ref{eq:ratio_ch_mom_abs}) whereas others remain exactly as for the total negativity. This provides a useful hint as to how more complicated formulae for multiparticle states will generalise to the symmetry resolved measure, namely by the substitutions $r_A \mapsto e^{\frac{2\pi i \alpha}{n}} r_A$ and $r_B \mapsto e^{-\frac{2\pi i \alpha}{n}} r_B$. Therefore, we can write:
\be
\label{phase_ansatz}
\mathcal{R}_1^n(r_A, r_B, r; \alpha) = \mathcal{R}_1^n\l e^{\frac{2\pi i \alpha}{n}} r_A, e^{-\frac{2\pi i \alpha}{n} }r_B, r; 0\r\,.
\ee

Indeed, the generalisation to states of many distinct quasi-particles is straightforward, and each particle contributes independently (multiplicatively) to the ratio of charged moments, similar to the structure found in \cite{partI,partII} for the charged R\'enyi entropies. We highlight that the results \eqref{eq:ratio_ch_mom_abs} are identical for fermions with the definitions \eqref{Rtrace} and \eqref{fermion_ch_moments_def} when $n$ is even. This is the case we will consider in the following when treating the fermionic case. For bosonic systems, states of multiple identical excitations can also be considered, whose total negativity was obtained in \cite{castro2019entanglement,graphpaper}. A similar formula can be derived for the ratio of moments, giving 
\beq
\mathcal{R}_k^n(r_A,r_B, r; \alpha)=\sum_{p=-k}^k   \sum_{q=\max(0,-np)}^{{[\frac{n}{2}(k-p)]}}\mathcal{A}_{p,q}  r_A^{n p+q} r^{n(k-p)-2q} r_B^{q} \, e^{2\pi i \alpha p},\,
\label{manyk}
\eeq
where $[\cdot]$ represents the integer part and 
\beq
\mathcal{A}_{p,q}= \sum_{\{k_1,\ldots,k_n \} \in \sigma_0^n(q)} \prod_{j=1}^n \frac{k!}{(p+k_j)! (k-p-k_j-k_{j+1})! k_j!}
\label{apq}
\eeq
are combinatorial factors, with the sum running over all the partitions $\{k_1,\ldots,k_n\}$ of the number $q$ into $n$ non-negative integer parts. 

\medskip 
We structure the paper as follows. In Section \ref{sec:Qubit} we analyse in detail a simplified model consisting of a state of few qubits. In spite of the simplicity of these states, their symmetry resolved negativity moments capture the main universal features found in QFT. In Section \ref{sec:Twist_op} we give a field theoretical formulation of the charged moments of the partially transposed density matrix, employing the notion of twist operators. The difference between fermionic and bosonic particles is 
thoroughly discussed again in this context, and the evaluation of the moments is shown to reduce to a combinatorial problem which we solve exactly for single and multiple distinct excitations. We check numerically our predictions on a 1D Fermi chain in Section \ref{sec:Numerics} and find good agreement. We conclude in Section \ref{conclusion}.

\section{Qubit Computation}\label{sec:Qubit}
In this section we derive the main formulae for the charged replica negativities by starting with a multi-qubit system. This ``toy model" was already employed in \cite{PRLexcited,castro2018entanglement,castro2019entanglement, partI, partII} following the realisation that even if multi-qubit states are much simpler than the excited states of a QFT, they both produce the same universal contribution to entanglement entropies and negativities. Hence it is advantageous to obtain such contribution from qubit states rather than a vastly more involved field-theoretical approach. 

The advantage of this picture is that the ground state of a multi-qubit system is trivial from the point of view of the entanglement content, which means that what is dubbed ``excess of entropy" or ``excess of negativity" or, in our case, ``ratio of moments" of an excited state with respect to the ground state effectively reduces to the entropy or negativity of the excited state and the moments of the excited state, respectively. The notion of charge imbalance in a qubit setup was introduced in \cite{cornfeld2018imbalance} and the notion of fermionic partial transposition in the same setup was later used in \cite{murciano2021symmetry}. 

The main result of this section is to show that equation \eqref{eq:ratio_ch_mom_abs} for a state consisting of a single excitation can be derived employing either the bosonic or the fermionic notion of partial transposition. Interestingly, even if the intermediate steps of the computation are different in the two cases, the final result is still the same . For bosonic theories, the result can be generalised to multiple identical excitations to give (\ref{manyk}) with (\ref{apq}).
\subsection{Single Bosonic Excitation}
 Assume that a single bosonic excitation is localised in space according to a uniform probability distribution, so that $r_A$, $r_B$ and $r$ can be regarded as the probabilities for the excitation to be found in regions $A$, $B$, $C$ respectively. Then the state in $\mathcal{H}_{A\cup B \cup C}$ representing a single excitation can be written as
\be
\label{qubit single particle state}
|{\boldsymbol{1}} \ket = \sqrt{r_A}|100\ket + \sqrt{r_B}|010\ket + \sqrt{r}|001\ket \,
\ee
where the values $0\,(1)$ represent the absence (presence) of the excitation and the coefficients can be interpreted as probabilities of finding the excitation in a particular region. 
The RDM $\rho_{A \cup B}$ is obtained taking the trace over $\mathcal{H}_C$:
\begin{align}
\rho_{A \cup B} = \Tr_{C} |{\boldsymbol{1}}\ket\bra {\boldsymbol{1}}| = r_A |10\ket\bra 10 | + r_B |01\ket \bra 01 | + \sqrt{r_A r_B}(|01\ket \bra 10 | + |10\ket \bra 01 |) + r|00\ket \bra 00 |\,,
\end{align}
or in matrix form
\be
\rho_{A \cup B} = 
\l
\begin{array}{c|cccc}
&00 & 01 & 10 & 11\\
\hline
00 & r & 0 & 0 & 0\\
01 & 0 & r_B & \sqrt{r_A r_B} & 0\\
10 & 0 & \sqrt{r_A r_B} & r_A & 0\\
11 & 0 & 0 & 0 & 0
\end{array}
\r\,.
\ee
This matrix has a block-diagonal structure with respect to the number operator in $A\cup B$:
\be
(N_A + N_B) |i_A,j_B\ket = (i_A + j_B)|i_A,j_B\ket\,, \quad i_A, j_B \,\in \{0,1\}
\ee
as indeed we can decompose it as
\be
\rho_{A \cup B} = (r)_{N=0} \oplus 
\begin{pmatrix}
     r_B & \sqrt{r_A r_B}\\
     \sqrt{r_A r_B} & r_A
  \end{pmatrix}_{N=1} \oplus (0)_{N=2}\, ,
\ee
where each block corresponds to an eigenspace of $N\equiv N_A+N_B$.
Let us now come to the partially transposed matrix $\rho_{A \cup B}^{T_B}$. From the definition \eqref{eq:bosonic_transp}, it follows:
\begin{align}
\rho_{A \cup B}^{T_B} &= r_A |10\ket\bra 10 | + r_B |01\ket \bra 01 | + \sqrt{r_A r_B}(|00\ket \bra 11 | + |11\ket \bra 00 |) + r|00\ket \bra 00 | \nonumber \\
&= 
\l
\begin{array}{c|cccc}
&00 & 01 & 10 & 11\\
\hline
00 & r & 0 & 0 & \sqrt{r_A r_B}\\
01 & 0 & r_B & 0 & 0\\
10 & 0 & 0 & r_A & 0\\
11 & \sqrt{r_A r_B} & 0 & 0 & 0
\end{array}
\r
\end{align}
which is a block-diagonal matrix with respect to the imbalance operator $\Delta N \equiv N_A-N_B$:
\be
(N_A - N_B) |i_A,j_B\ket = (i_A - j_B)|i_A,j_B\ket\,, \quad i_A, j_B \,\in \{0,1\}\,,
\ee
as
\be
\label{eq:qubit_boson_partial_transpose}
\rho_{A \cup B}^{T_B} = (r_A)_{\Delta N=1} \oplus 
\begin{pmatrix}
     r & \sqrt{r_A r_B}\\
     \sqrt{r_A r_B} & 0
  \end{pmatrix}_{\Delta N=0} \oplus (r_B)_{\Delta N=-1}\,.
\ee
The spectrum of $\rho_{A \cup B}^{T_B}$ contains four distinct eigenvalues, one of which is negative and produced by the block $\Delta N=0$. The eigenvalues are
\be
 \left \{r_A,\, r_B,\, \frac{r+\sqrt{r^2 + 4r_Ar_B}}{2},\,\frac{r-\sqrt{r^2 + 4r_Ar_B}}{2}\right\}
\ee
and therefore this system has non-vanishing negativity. The block-diagonal structure of $\rho_{A \cup B}^{T_B}$, i.e. the property $[\rho_{A \cup B}^{T_B},\, N_A - N_B]=0$, implies that the operator $e^{2\pi i \alpha (N_A - N_B)}$ attaches a phase $e^{2\pi i \alpha}$ to the $\Delta N=1$ block, a phase $e^{-2\pi i \alpha}$ to the $\Delta N=-1$ block and acts as the identity on the uncharged block. Hence, we finally obtain the expected result:
\be
\label{eq:ratio_ch_mom_qubit_single}
\text{Tr}\l (\rho^{T_B}_{A\cup B})^n e^{2\pi i \alpha (N_A-N_B)}\r = \mathcal{R}_1^n(r_A, r_B, r;\alpha)\,,
\ee
with $\mathcal{R}_1^n(r_A, r_B, r;\alpha)$ given by (\ref{eq:ratio_ch_mom_abs}). In addition, knowing the eigenvalues allows to compute the negativity from its original definition \eqref{eq:def_negativity}, without the need to obtain the moments first. 

\subsection{Single fermionic excitation}
In this section we will show that the result \eqref{eq:ratio_ch_mom_abs} holds for $n$ even also if we adopt the fermionic definition of partial transposition \eqref{eq:ferm_transp}. It is important to emphasise beforehand that by speaking of fermionic excitations in this context we only refer to the prescription for the partial transposition of the RDM, not to any algebra of the operators that create and annihilate the qubit states.

If we adopt the definition \eqref{eq:ferm_transp}, the matrix $\rho^{R_B}_{A\cup B}$ differs from $\rho^{T_B}_{A\cup B}$ as obtained in the previous section only because there is now an extra phase in the off-diagonal elements:
\be
(|10\ket \bra 01|)^{R_B} = -i |11\ket \bra 00|\,, \quad (|01\ket \bra 10|)^{R_B} = -i |00\ket \bra 11|
\ee
while the diagonal elements are not modified. It follows that $\rho^{R_B}_{A\cup B}$ is still block-diagonal with respect to the imbalance operator:
\be
\rho_{A \cup B}^{R_B} = (r_A)_{\Delta N=1} \oplus 
\begin{pmatrix}
     r & -i\sqrt{r_A r_B}\\
     -i\sqrt{r_A r_B} & 0
  \end{pmatrix}_{\Delta N=0} \oplus (r_B)_{\Delta N=-1}\,.
\ee
The eigenvalues of the zero charge sector can now be imaginary, depending on the values of $r_A$ and $r_B$. However, we are eventually interested in the evaluation of the charged moment $\text{Tr}\l |\rho^{R_B}_{A\cup B}|^n e^{i2\pi \alpha (N_A-N_B)}\r$, which requires the knowledge of the eigenvalues of $|\rho^{R_B}_{A\cup B}|$ only. From the definition(\ref{Rtrace}), and making use of the block diagonal decomposition, it is clear that we need to find the spectrum of the matrix:
\be
\begin{pmatrix}
     r & -i\sqrt{r_A r_B}\\
     -i\sqrt{r_A r_B} & 0
\end{pmatrix}
\begin{pmatrix}
     r & -i\sqrt{r_A r_B}\\
     -i\sqrt{r_A r_B} & 0
\end{pmatrix}^\dag 
= 
\begin{pmatrix}
     r^2 + r_Ar_B & ir\sqrt{r_A r_B}\\
     -ir\sqrt{r_A r_B} & r_Ar_B\,,
\end{pmatrix}
\ee
which is given by
\be
\lambda_\pm = \l \frac{r \pm \sqrt{r^2 + 4r_Ar_B}}{2} \r^2\,.
\ee
The eigenvalues $\lambda_\pm$ are nothing but the squares of the eigenvalues of $|\rho_{A \cup B}^{R_B}|$ in the sector with $\Delta N = 0$, while the ones associated to $\Delta N = \pm 1$ are given by $r_A,r_B$ respectively.
Since this spectrum was already obtained in the previous section for a bosonic particle,the result \eqref{eq:ratio_ch_mom_qubit_single} is recovered here for $n$ even.

\subsection{Multiple Distinct Excitations}
 A generic state consisting of $k$ distinct excitations is a linear combination of states of the form
\be
|i_A^1\dots i_A^k\,,i_B^1\dots i_B^k\,,i_C^1\dots i_C^k\ket \equiv \otimes_{j=1}^k|i_A^j,i_B^j,i_C^j\ket \, \in \, \l \mathcal{H}_{A \cup B \cup C} \r^{\otimes k} 
\ee
and for every $j=1,\dots,k$
\be
i^j_A, i^j_B, i^j_C \, \in \{0,1\} \,, \quad i^j_A+ i^j_B+ i^j_C = 1 \,.
\ee
Among these states, let us focus on those which are tensor products of the linear combination \eqref{qubit single particle state}, that is, on states of the form $|{\boldsymbol{1}}{\boldsymbol{1}}\ldots {\boldsymbol{1}}\ket := |{\boldsymbol{1}}\ket^{\otimes k}$. For $k=2$ we have for instance
\begin{align}
|{\boldsymbol{1}}{\boldsymbol{1}} \ket &= |{\boldsymbol{1}}\ket^{\otimes 2} = r_A |11,00,00\ket + r_B |00,11,00\ket + r|00,00,11\ket \nonumber \\
&+\sqrt{r_A r_B} (|10,01,00\ket +  |01,10,00\ket) \nonumber \\ &+\sqrt{r_A r} (|10,00,01\ket +  |01,00,10\ket)\nonumber \\ &+\sqrt{r_B r} (|00,10,01\ket + |00,01,10\ket) \,.
\end{align}
For a tensor product state the density matrix is the tensor product of the single-particle density matrices, and since the trace of a tensor product is the product of the traces, the RDM is a tensor product itself:
\be
\rho^{({\boldsymbol{1}},\dots,{\boldsymbol{1}})} = (\rho^{({\boldsymbol{1}})})^{\otimes k}\,, \quad \rho_{A\cup B}^{({\boldsymbol{1}},\dots,{\boldsymbol{1}})} = \text{Tr}_C\rho^{({\boldsymbol{1}},\dots,{\boldsymbol{1}})} = (\rho_{A\cup B}^{({\boldsymbol{1}})})^{\otimes k}\,.
\ee
Now we can compute the charged replica negativities using the bosonic partial transposition \eqref{eq:bosonic_transp}:
\begin{align}
&\text{Tr}\l \l\rho^{({\boldsymbol{1}},\dots,{\boldsymbol{1}})\,,T_B}_{A\cup B}\r^n e^{2\pi i \alpha (N_A-N_B)}\r = \text{Tr}\l \l(\rho^{({\boldsymbol{1}})\,,T_B}_{A\cup B})^{\otimes k}\r^n e^{2\pi i \alpha (N_A-N_B)}\r \nonumber \\
= &\text{Tr}\l \l\l\rho^{({\boldsymbol{1}})\,,T_B}_{A\cup B}\r^n\r^{\otimes k} e^{2\pi i \alpha (N_A-N_B)}\r =\text{Tr}\l \l\rho^{({\boldsymbol{1}})\,,T_B}_{A\cup B}\r^n e^{2\pi i\alpha (N^{({\boldsymbol{1}})}_A-N^{({\boldsymbol{1}})}_B)}\r^{\otimes k}\nonumber \\
= & \l \mathcal{R}_1^n(r_A,r_B,r;\alpha)\r^k\,,
\end{align}
where the operators $N_A^{(\boldsymbol{1})}$, $N_B^{(\boldsymbol{1})}$ act on a single two-qubit state in the obvious way 
\be
N^{(\boldsymbol{1})}_A|i_A,i_B\ket = i_A |i_A,i_B\ket\,, \quad N^{(\boldsymbol{1})}_B|i_A,i_B\ket = i_B |i_A,i_B\ket\,, \quad i_A\,,i_B\,\in \{0,1\}\,.
\ee
The result above is not surprising and it is a consequence of the choice of the state: if the multi-particle state is a tensor product then there is no correlation between different particles and the total negativity is simply the product of the single particle negativities. This also holds for the ratio of charged moments. 
As we shall see below, this is not the case when the particles are indistinguishable.

\subsection{Multiple Identical Excitations}
Consider now a $k$-particle state consisting of $k$ identical excitations. Its associated qubit state can be written as:
\beq
 |\boldsymbol{k}\ket = \sum_{\{k_A,k_B, k_C\} \in \sigma^3_0(k)} c_{k_A,k_B, k_C} \,|k_A k_B k_C \ket\,,
\eeq
where $\sigma^3_0(k)$ represents the set of integer partitions of $k$ into three non-negative parts and the coefficients
\beq
c_{k_A,k_B, k_C}:=\sqrt{\frac{k! r_A^{k_A} r^{k_C} r_B^{k_B}}{k_A! k_B! k_C!}} \delta_{k_A+k_B+k_C,k}\,,
\label{cdef}
\eeq
are as usual probabilities of finding $k_A$ identical particles in region $A$, $k_B$ identical particles in region $B$ and the remaining particles, $k_C$ in region $C$, weighted with the appropriate combinatorial factors. As shown in \cite{castro2019entanglement,graphpaper}, if all vectors $|k_A k_B k_C\ket$ are normalised to one, then also $ \bra \boldsymbol{k}|\boldsymbol{k}\ket=1$. 

From this expression it is then possible to explicitly construct the matrix elements of the (bosonic) partially transposed density matrix as:
\beqa
\bra k_A^1 k_B^1|\rho^{T_B}_{A \cup B}|k^2_A k^2_B\ket = \sum_{k_C \in \mathbb{N}_0}  
c_{k_A^1 k^2_B k_C} c_{k^2_A k^1_B k_C}\,,
\label{43}
\eeqa
where the sum represents taking the trace over the degrees of freedom in $C$ and the partial transposition exchanges the indices $k^1_B$ and $k^2_B$ in the coefficients.
The matrix elements of the $n$th power can then be simplified to 
\beqa
\bra k^{1}_A k^{1}_B|\left(\rho^{T_B}_{A \cup B}\right)^n|k^{n+1}_A k^{n+1}_B\ket&=&  \sum_{k_A^s, k_B^s \in \N_0; s=2,\ldots,n \atop k_C^r \in \N_0; r \in I_n}	 \prod_{j=1}^n c_{k^{j}_A k_B^{j+1} k_C^{j}} c_{k^{j+1}_A k^{j}_B k^{j}_C}\\
&=&\sum_{k_A^s, k_B^s \in \N_0; s=2,\ldots,n \atop k_C^r \in \N_0; r \in I_n}\prod_{j=1}^n \frac{k!  r_A^{k_A^{j}} r^{k_C^{j}}r_B^{k_B^{j}}}{k_A^{j}! k^{j}_C! k_B^{j}!}  \delta_{k_A^{j}+k_B^{j+1}+k_C^{j},k} \, \delta_{k_A^{j+1}+k_B^{j}+k_C^{j},k} \,,\nonumber
\label{trace2}
\eeqa
where $I_n:=\{1,\ldots, n\}$. This formula follows from multiplying together $n$ copies of (\ref{43}) with different intermediate states (indices). 

While Eq. (\ref{trace2}) was already presented in \cite{castro2019entanglement}, its symmetry resolved version is new and can be easily written by introducing phase factors in the sum above. We have 
\beqa
&& \bra k^{1}_A k^{1}_B|\left(\rho^{T_B}_{A \cup B}\right)^n e^{2\pi i \alpha (N_A-N_B)}|k^{n+1}_A k^{n+1}_B\ket\nonumber\\
&& \quad=\sum_{k_A^s, k_B^s \in \N_0; s=2,\ldots,n \atop k_C^r \in \N_0; r \in I_n}\prod_{j=1}^n \frac{k!  r_A^{k_A^{j}} r^{k_C^{j}}r_B^{k_B^{j}}}{k_A^{j}! k^{j}_C! k_B^{j}!}  \, e^{\frac{2\pi i \alpha}{{n}}(k_A^j-k_B^j)}\delta_{k_A^{j}+k_B^{j+1}+k_C^{j},k} \, \delta_{k_A^{j+1}+k_B^{j}+k_C^{j},k} \,.
\label{trace3}
\eeqa
Starting with this result, the derivation of equations (\ref{manyk}) and (\ref{apq}) is identical to that presented in \cite{castro2019entanglement,graphpaper}. The idea is to employ the various existing constraints on the values of $k^j_A, k^j_B$ and $k^j_C$ in order to reduce the number of terms in the sum and product (\ref{trace3}). Several terms in the sum are vanishing due to the two delta-function constraints. Once those are implemented, only one independent set of variables, say $k_A^j$ remains. This set itself is constrained by the fact that each of these numbers can never be larger than $k$. The implementation of all these constraints eventually leads to the result (\ref{manyk}) with (\ref{apq}). Before concluding the section let us analyse the simplest case, $k=1$.
    For a single excitation, the right-hand side of \eqref{manyk} is given by:
    \beq
    \mathcal{A}_{-1,n}\,r_B^n\, e^{-2\pi i \alpha} + \mathcal{A}_{1,0}\,r_A^n\, e^{2\pi i \alpha} +\sum_{q=0}^{[n/2]}\mathcal{A}_{0,q}(r_A r_B)^qr^{n-2q} \nonumber\,.
    \eeq
    By looking at the definition \eqref{apq} one immediately gets $\mathcal{A}_{-1,n}= \mathcal{A}_{1,0}=1$. On the other hand, $\mathcal{A}_{0,q}=\sum_{\{k_1,\ldots,k_n \}\in \sigma_0^n(q)} 1$, where each $k_j\,\in\{0,1\}$ and whenever $k_j=1$ then $k_{j+1}=0$. Counting the number of sequences $(k_1,\,\dots\,,k_n)$ that satisfy these constraints is a combinatorial problem identical to the one we solve in the next section. This number is $\frac{n}{n-q}\binom{n-q}{n}$. As we explain in the next section, this result exactly reproduces \eqref{eq:ratio_ch_mom_abs}.

Looking at the coefficients (\ref{apq}) it is clear that the computation has an underlying combinatorial interpretation. For the $\alpha=0$ case this has been established by reinterpreting the sum (\ref{manyk}) as a partition function for a certain class of graphs \cite{graphpaper}. A combinatorial picture will emerge again in the next section in a related context: the computation using twist operators. 

\section{Twist operator approach}\label{sec:Twist_op}
In this section, we provide a field theoretic description of the charged moments of the partially transposed RDM, valid in principle for any QFT. To do so, we employ the replica construction, and we define a set of {\it twist operators}, which are based on the branch point twist fields of 1+1D QFT \cite{Calabrese:2004eu,entropy,GS,SymResFF}. Branch point twist fields play a prominent role in entanglement computations in 1+1D QFT and were used also in the context of excited states in \cite{PRLexcited,castro2018entanglement,castro2019entanglement,graphpaper,partI}. Branch point twist fields sit at branch points from which branch cuts extend. Twist operators were introduced in a previous work \cite{partII}, and have very similar exchange relations with respect to local fields as branch point twist fields, with the difference that they act on extended regions of space rather than points.

In \cite{partII} we showed that the analysis of the entanglement entropies of quasi-particle states relies on few algebraic properties, mainly their exchange relations w.r.t.~local fields, while most of the theory-dependent features are hidden in the zero-point fluctuations (ground-state entanglement). 
Similar to the previous section we will consider here a single excitation of either bosonic or fermionic type. 
In the fermionic case, we discuss the algebra of twist operators and fermionic fields. We both compute the charged moments in a generic fermionic theory, and, as a byproduct, we perform a simpler derivation valid for free fermions in any dimension.

\subsection{Single Bosonic Excitation}

We consider now a bosonic QFT, described by its algebra of observables $\mathcal{A}$, acting on the Hilbert space $\mathcal{H}$, and $|0\ket \in \mathcal{H}$ is the ground-state of the theory. We then consider the replica version of this theory, consisting of $n$ non-interacting copies of the same model. The algebra of observables is now denoted by $\mathcal{A}^n$, so that $\mathbb{Z}_n$ becomes an internal (global) symmetry, which includes cyclic permutation symmetry among copies \cite{kniz,orbifold,Bouwknegt,Borisov}. We also assume that the QFT we start with carries an additional global $U(1)$ symmetry. This procedure allows us to introduce a set of twist operators, supported on extended spacial regions, which mix cyclic permutation and internal $U(1)$ symmetries, and generalise the notion of composite twist fields of 1+1D QFT \cite{ctheorem,Levi,BCDLR,FBoson,GS,SymResFF}.\\

Let $\mathcal{O}^j(\mathbf{x})$ be a generic bosonic field of the $j$-th replica ($j=1,\dots,n$) with $U(1)$ charge $\kappa_\mathcal{O}$. We consider a region $A$, and we associate to it a twist operator $T^{\alpha}_A$ which satisfies \cite{partII}
\be\label{eq:bos_twist}
T^\alpha_A \mathcal{O}^j(\mathbf{x}) = \begin{cases} e^{2\pi i \kappa_\mathcal{O} \alpha \delta_{j,n}} \mathcal{O}^{j+1}(\mathbf{x}) T^\alpha_A \quad \mathbf{x} \in A\,, \\
\qquad \qquad \quad \,\mathcal{O}^{j}(\mathbf{x}) T^\alpha_A \quad \mathbf{x} \notin A\,.
\end{cases}
\ee
The action of $T^\alpha_A$ is non-trivial only in the region $A$ and it can be interpreted as a replica shift $j\rightarrow j+1$ followed by the insertion of a $U(1)$ flux among the $n$-th and the first replica.
Similarly, we define the conjugate twist operator $\tilde{T}^{\alpha}_A$ so that
\be
\tilde{T}^\alpha_A \mathcal{O}^j(\mathbf{x}) = \begin{cases} e^{-2\pi i \kappa_\mathcal{O} \alpha \delta_{j,1}} \mathcal{O}^{j-1}(\mathbf{x}) \tilde{T}^\alpha_A \quad \mathbf{x} \in A\,, \\
\qquad \qquad \quad \,\mathcal{O}^{j}(\mathbf{x}) \tilde{T}^\alpha_A \quad \mathbf{x} \notin \,A,
\end{cases}
\ee
whose action is the inverse of that of $T^\alpha_A$, so we can identify $\tilde{T}^\alpha_A = (T^\alpha_A)^\dagger$. These operators will now allow us to develop a field-theoretic formulation of the symmetry-resolved negativity and its moments. Let $A$ and $B$ be two disconnected regions, $|\Psi\ket$ a state and $\rho_{A \cup B}$ its RDM over $A \cup B$. Following \cite{ssr-17}, one can indeed interpret the moments as charged functions of a $n$-sheeted Riemann surface with a branch-cut over $A$ and $B$, connecting the replicas in two opposite directions. A similar construction in the presence of fluxes has been proposed in \cite{murciano2021symmetry}. In analogy with the works above, we establish the following relation between the charged moments and the twist operators:
\be
\text{Tr}\l (\rho^{T_B}_{A\cup B})^{n} e^{i2\pi \alpha (Q_A-Q_B)}\r \sim {}^{n}\bra \Psi | T^\alpha_A \tilde{T}^\alpha_B| \Psi \ket^n\,,
\ee
up to a non-universal proportionality constant, which is irrelevant for our purpose as we are interested in ratios. Note that $|\Psi\ket^n$ represents the replicated version of the state $|\Psi\ket$.

Consider once more the simplest case of a one-particle excitation of fixed momentum $\mathbf{p}$ and charge $ 1$, which is created by a field $\mathcal{O}$ acting on the ground-state as 
\be
|\boldsymbol{1}\ket \propto \mathcal{O}(\mathbf{p})| 0\ket\,,
\ee
up to a normalization constant. Here $\mathcal{O}(\mathbf{p})$ is the Fourier transform of $\mathcal{O}(\mathbf{x})$
\be
\mathcal{O}(\mathbf{p}) = \int_{\mathcal{M}} d^dx e^{-i\mathbf{p\cdot x}}\,\mathcal{O}(\mathbf{x})\,,
\ee
and $\mathcal{M}$ is the whole space, which, for simplicity, we take to be a $d$-dimensional torus.
We note that  $\mathcal{O}^\dagger(-\mathbf{p})=[\mathcal{O}(\mathbf{p})]^\dagger$ which will be important for later computations.
Our aim, as in the previous section, is to compute the ratio of charged moments, which in this language becomes
\be\label{eq:ratio_ch_moments}
 \frac{{}^{n}\bra \boldsymbol{1} | T^\alpha_A \tilde{T}^\alpha_B| \boldsymbol{1} \ket^n}{{}^{n}\bra 0 | T^\alpha_A \tilde{T}^\alpha_B| 0 \ket^n}\,.
\ee
  We define the projection of $\mathcal{O}(\mathbf{p})$ over a generic region $A$ as the restricted integral
\be
\mathcal{O}_A(\mathbf{p}) = \int_{A}d^dx e^{-i\mathbf{p \cdot x}}\mathcal{O}(\mathbf{x})\,,
\ee
then we can write 
\be
\begin{split}
{}^{n}\bra \boldsymbol{1} | T^\alpha_A \tilde{T}^\alpha_B| \boldsymbol{1} \ket^n = \frac{^n\bra 0 |(\mathcal{O}^\dagger)^n(-\mathbf{p})\dots(\mathcal{O}^\dagger)^1(-\mathbf{p}) T^\alpha_A \tilde{T}^\alpha_B \mathcal{O}^1(\mathbf{p})\dots \mathcal{O}^n(\mathbf{p})|0\ket^n}{^n\bra 0 |(\mathcal{O}^\dagger)^n(-\mathbf{p})\dots(\mathcal{O}^\dagger)^1(-\mathbf{p}) \mathcal{O}^1(\mathbf{p})\dots \mathcal{O}^n(\mathbf{p})|0\ket^n}\,. \label{ratio}
\end{split}
\ee
We point out that bosonic creation operators, whether on the same or on different copies, commute with each other, therefore the order of a string of operators $\mathcal{O}^j(\boldsymbol{p})$ is irrelevant. We now observe that
\be
\mathcal{O}^j(\mathbf{p}) = \mathcal{O}^j_A(\mathbf{p}) + \mathcal{O}^j_B(\mathbf{p})  + \mathcal{O}^j_{C}(\mathbf{p})\,,
\ee
which, when inserted into (\ref{ratio}), leads to a large numbers of terms both in the numerator and the denominator. The key idea is that in the infinite volume limit many of these terms are subleading and the leading contribution can be isolated and shown to be simple. We now present the details of the calculation. 

Employing the exchange relations between twist operators and $\mathcal{O}$, we can bring all the bosonic fields $\mathcal{O}^j(\mathbf{p})$ to the left of $T^\alpha_A \tilde{T}^\alpha_B$ in (\ref{ratio}). This gives
\be
\begin{split}
^n\bra 0 |(\mathcal{O}^\dagger)^n(-\mathbf{p})\dots(\mathcal{O}^\dagger)^1(-\mathbf{p}) T^\alpha_A \tilde{T}^\alpha_B \mathcal{O}^1(\mathbf{p})\dots \mathcal{O}^n(\mathbf{p})|0\ket^n = \\
^n\bra 0 |(\mathcal{O}^\dagger)^n(-\mathbf{p})\dots(\mathcal{O}^\dagger)^1(-\mathbf{p})  (\mathcal{O}^2_A(\mathbf{p})+\mathcal{O}^n_B(\mathbf{p})e^{-2\pi i \alpha}+\mathcal{O}^1_{C}(\mathbf{p}))\dots \times\\
(\mathcal{O}^1_{A}(\mathbf{p})e^{2\pi i \alpha} + \mathcal{O}^{n-1}_{B}(\mathbf{p})+ \mathcal{O}^n_{C}(\mathbf{p}))T^\alpha_A \tilde{T}^\alpha_B|0\ket^n\,.
\end{split}
\ee
So far everything is exact, and no approximation has been made. To proceed further with the evaluation of the expectation value, we focus on the large volume behavior. In \cite{partII} we argued that the leading terms come from the contractions of fields belonging to the same replica, which amount to the following formal replacement inside the correlation function:
\be
(\mathcal{O}^\dagger)^j_{A'}(-\mathbf{p})\mathcal{O}^j_{A}(\mathbf{p}) \rightarrow \bra 0 |(\mathcal{O}^\dagger)^j_{A'}(-\mathbf{p})\mathcal{O}^j_{A}(\mathbf{p}) |0\ket \sim V_{A\cap A'}\,,
\label{55}
\ee
with $A,\,A' \subseteq \mathcal{M}$ generic spacial regions. Proportionality to the volume is valid in the large volume limit and was shown in \cite{partII}. The proportionality constant can in principle be absorbed in the normalisation of the field $\mathcal{O}$ and does not affect the final result. In this regime, the vacuum expectation value of the twist operators also factors out: 
\be\label{eq:manyTerms}
\begin{split}
^n\bra 0 |(\mathcal{O}^\dagger)^n(-\mathbf{p})\dots(\mathcal{O}^\dagger)^1(-\mathbf{p})  (\mathcal{O}^2_A(\mathbf{p})+\mathcal{O}^n_B(\mathbf{p})e^{-i2\pi \alpha}+\mathcal{O}^1_{C}(\mathbf{p}))\dots \times\\
(\mathcal{O}^1_{A}(\mathbf{p})e^{i2\pi \alpha} + \mathcal{O}^{n-1}_{B}+ \mathcal{O}^n_{C}(\mathbf{p}))T^\alpha_A \tilde{T}^\alpha_B|0\ket^n \simeq \\
^n\bra 0 |(\mathcal{O}^\dagger)^n(-\mathbf{p})\dots(\mathcal{O}^\dagger)^1(-\mathbf{p})  (\mathcal{O}^2_A(\mathbf{p})+\mathcal{O}^n_B(\mathbf{p})e^{-2\pi i \alpha}+\mathcal{O}^1_{C}(\mathbf{p}))\dots \times\\
(\mathcal{O}^1_{A}(\mathbf{p})e^{2\pi i \alpha} + \mathcal{O}^{n-1}_{B}(\mathbf{p})+ \mathcal{O}^n_{C}(\mathbf{p}))|0\ket^{n} \times {}^{n}\bra 0 |T^\alpha_A \tilde{T}^\alpha_B|0\ket^n\,,
\end{split}
\ee
which means that the charge moments of the ground state factor out and will subsequently be cancelled in the ratio \eqref{eq:ratio_ch_moments}. While the formula above is already a large volume approximation, when the sums are expanded and each individual term considered, many terms that are subleading for large volume still appear. To make things clear, consider the terms
\be\label{eq:string_A}
^n\bra 0 |(\mathcal{O}^\dagger)^n(-\mathbf{p})\dots(\mathcal{O}^\dagger)^1(-\mathbf{p})\mathcal{O}^2_A(\mathbf{p})\dots \mathcal{O}^1_A(\mathbf{p})|0\ket^n e^{2\pi i\alpha} \sim V^n_A e^{2\pi i\alpha}\,,
\ee
and
\be\label{eq:string_B}
^n\bra 0 |(\mathcal{O}^\dagger)^n(-\mathbf{p})\dots(\mathcal{O}^\dagger)^1(-\mathbf{p})\mathcal{O}^n_B(\mathbf{p})\dots \mathcal{O}^{n-1}_B(\mathbf{p})|0\ket^n e^{-2\pi i\alpha} \sim V^n_B e^{-2\pi i\alpha}\,.
\ee
These are the terms that generate the highest powers in the volume of regions $A$ and $B$ respectively. Among the other terms that are generated, the leading ones at large volume are those containing a string of operators $\mathcal{O}^j(\mathbf{p})$ and their daggered versions all inserted at different replicas, just as in the examples above. If that is not the case, there is at least a pair of operators that can not be contracted as in \eqref{55} and the term is subleading.

We now proceed with the systematic evaluation and counting of all the leading terms generated in the expansion \eqref{eq:manyTerms}. We introduce the following notation to identify each term
\be\label{eq:string_notation}
(A_1 \dots  A_n):=\,
^n\bra 0 |(\mathcal{O}^\dagger)^n(-\mathbf{p})\dots(\mathcal{O}^\dagger)^1(-\mathbf{p})\mathcal{O}^{j_1}_{A_1}(\mathbf{p})\dots \mathcal{O}^{j_n}_{A_n}(\mathbf{p})|0\ket^n,
\ee
where $A_i \in \{A,B,C\}$ and $j_i \in \{1,\dots,n\}$. 
We observe that, once the sequence of regions $(A_1 \dots  A_n)$ is identified, the sequence of replica indices $(j_1  \dots  j_n)$ is fixed unambiguously, as in fact
\be
\label{strings_J_i_values}
j_i = 
\begin{cases}
i+1\,,\quad &\text{if }A_i = A \\
i\,,\quad &\text{if }A_i = C \\
i-1\,,\quad &\text{if }A_i = B
\end{cases}\quad ,
\ee
hence the choice of notation above. Moreover, due to the contraction rules discussed above, only the terms for which $(j_1 \dots  j_n)$ is a permutation of the indices $\{1,\dots,n\}$ are non-vanishing. As a consequence, one can show that (See Appendix \ref{sec:nonzero_strings}) the only possible non-vanishing terms fit into one of these two categories:
\begin{itemize}
    \item Either $(A_1 \dots  A_n) = (A\dots A)$ or $(A_1 \dots  A_n) = (B\dots B)$, the two cases which have been already discussed in Eqs.  \eqref{eq:string_A}, \eqref{eq:string_B},
    \item Or, whenever $A$ appears in $(A_1 \dots  A_n)$, it has to be followed by $B$. Similarly, if $B$ appears in $(A_1 \dots  A_n)$, then it has to be preceded by $A$.
\end{itemize}
We focus on the second set of terms. It is convenient to split this into two additional subsets, which we call type-I and type-II
\begin{itemize}
    \item Type-I: $(A_1 \dots  A_n) = (B \ A_2 \dots A_{n-1} \ A )$\,,
    \item  Type-II: $(A_1 \dots  A_n)$ with $A_1\neq B$\,.
\end{itemize}
Thus, both types of string contain a number $k$ of pairs $AB$ and $n-2k$ C's, and according to (\ref{55}) each of them will be proportional to $(V_A V_B)^k V_C^{n-2k}$. Due to the balance of A's and B's there is no phase present in these terms (no $\alpha$ dependence). We now just need to count how many of each type we have. 

Among the strings of type-I, there is always at least one pair $AB$ and there are at most $k-1$ additional pairs AB that can be present. The number of such strings is precisely (a proof is given in \ref{sec:Comb_count})
\be
\binom{n-k-1}{k-1} \quad \mathrm{for} \quad k=0,\dots 
 [n/2].
\ee
Similarly, one can show that the number of type-II strings consisting of $k$ pairs of consecutive $A$ and $B$ is
\be
\binom{n-k}{k} \quad \mathrm{for} \quad k=0,\dots [n/2].
\ee
In summary,
\be\label{eq:all_contractions}
\begin{split}
^n\bra 0 |(\mathcal{O}^\dagger)^n(-\mathbf{p})\dots(\mathcal{O}^\dagger)^1(-\mathbf{p})  (\mathcal{O}^2_A(\mathbf{p})+\mathcal{O}^n_B(\mathbf{p})e^{-i2\pi \alpha}+\mathcal{O}^1_{C}(\mathbf{p}))\dots \times\\
(\mathcal{O}^1_{A}(\mathbf{p})e^{i2\pi \alpha} + \mathcal{O}^{n-1}_{B}(\mathbf{p})e^{i2\pi \alpha}+ \mathcal{O}^n_{C}(\mathbf{p}))|0\ket^{n} \sim\\
V^n_A e^{2\pi i\alpha} + V^n_B e^{-2\pi i \alpha} + \sum^{\lceil \frac{n}{2} \rceil}_{k=0}\frac{n}{n-k} \binom{n-k}{k} (V_AV_B)^{k}(V_{C})^{n-2k}\,,
\end{split}
\ee
where we used the simple identity
\beq 
\binom{n-k}{k}+ \binom{n-k-1}{k-1} =\frac{n}{n-k} \binom{n-k}{k}\,.
\eeq 
The denominator in \eqref{ratio} can be fully contracted and yields $V^n$ (up to a non-universal normalisation constant). Therefore the ratio \eqref{eq:ratio_ch_moments} becomes a function of the usual variables $r_A, r_B$ and $r$ and we we obtain 
\be
\label{eq:final_eq_bos}
\mathcal{R}_1^n(r_A,r_B,r;\alpha)=r^n_A e^{2\pi i \alpha} + r^n_B e^{-2\pi i\alpha}+\sum^{\lfloor \frac{n}{2} \rfloor}_{k=0}\frac{n}{n-k} \binom{n-k}{k} (r_Ar_B)^{k}r^{n-2k}\,,
\ee
which is the main result of this section. Note that although this formula looks different from (\ref{eq:ratio_ch_mom_abs}), they are in fact equivalent. That is
\be 
\label{Lucas_main}
\l \frac{r+\sqrt{r^2+4r_Ar_B}}{2}\r^{n} +\l \frac{r- \sqrt{r^2+4r_Ar_B}}{2}\r^{n} = \sum^{\lfloor \frac{n}{2} \rfloor}_{k=0}\frac{n}{n-k} \binom{n-k}{k} (r_Ar_B)^{k}r^{n-2k}\,.
\ee
This relation was given already in \cite{castro2019entanglement,graphpaper} without a proof. The proof is indeed quite involved, and can be performed using properties of the generalised Lucas' polynomials. This is presented in Appendix~\ref{Lucas}, where we also derive two interesting corollaries. The equality \eqref{Lucas_main} is particularly interesting because it shows that the result is always a polynomial in integer powers of $r_A, r_B, r$ for n positive, even or odd. However, its analytic continuation from $n$ even to $n=1$ does contain a square root as seen in (\ref{analy}).

\subsection{Single Fermionic Excitation}

Let us consider a theory for which the algebra $\mathcal{A}$ contains fermionic observables. In other words, we assume that $\mathcal{A}$ is a $\mathbb{Z}_2$-graded algebra (superalgebra) generated by bosonic/fermionic fields, which are even/odd with respect to the $\mathbb{Z}_2$ fermionic parity. Here, to generalise properly the twist operator construction, one must take care of the fermionic nature of the fields. Indeed, two such fields sitting at distinct points, say $\Psi(\mathbf{x})$ and $\Psi(\mathbf{x}')$ will now anticommute
\be
\Psi(\mathbf{x})\Psi(\mathbf{x}') = - \Psi(\mathbf{x}')\Psi(\mathbf{x}).
\ee
Moreover, when the replica construction is performed and the replica fields are obtained, we will require that fermionic fields on distinct replicas also anticommute
\be
\Psi^j(\mathbf{x})\Psi^{j'}(\mathbf{x}') = - \Psi^{j'}(\mathbf{x}')\Psi^j(\mathbf{x}).
\ee
As a result, the algebra of the replica theory $\mathcal{A}^n$ is not a conventional tensor product.

As before, we assume that an additional $U(1)$ symmetry is present in the theory. Let $A$ be a spacial region, and we associate to it a twist operator $T^{\alpha}_{A}$ which shifts the replica indices and appends a $U(1)$ flux. While its action for bosonic fields has been already discussed, we now consider its commutation relation with a generic fermionic field $\Psi$ of charge $\kappa_\Psi$. The natural generalisation is
\be\label{eq:ferm_twist}
T^\alpha_A \Psi^j(\mathbf{x}) = \begin{cases} (-1)^{(n-1)\delta_{j,n}}e^{2\pi i \kappa_\Psi \alpha \delta_{j,n}} \Psi^{j+1}(\mathbf{x}) T^\alpha_A \quad \mathbf{x} \in A, \\
\qquad \qquad \quad \,\Psi^{j}(\mathbf{x}) T^\alpha_A \quad \mathbf{x} \notin A.
\end{cases}
\ee
We point out that the only difference with respect to \eqref{eq:bos_twist} is the presence of an additional flux $(-1)^{n-1}$ between the $n$-th and the first replica, a factor that was already introduced in \cite{Casini} and employed for instance in \cite{entropy} in the calculation of the VEV of the Ising twist field.

For fermionic theories we need to define another twist operator which implements explicitly the fermionic partial transposition, and from now on we only consider $n$ even, denoted also by $n_e$. It has been shown in \cite{ssr-17} that the effect of the partial transposition on the fermions gives rise to an additional insertion of a flux $(-1)$ among any pair of consecutive replicas, in addition to the usual replica shift. To implement this construction, we define a twist operator $\tilde{T}^{\alpha}_A$ satisfying
\be
\tilde{T}^\alpha_A \Psi^j(\mathbf{x}) = \begin{cases} -(-1)^{(n-1)\delta_{j,1}}e^{-2\pi i \kappa_\Psi \alpha \delta_{j,1}} \Psi^{j-1}(\mathbf{x}) \tilde{T}^\alpha_A \quad \mathbf{x} \in A, \\
\qquad \qquad \quad \,\Psi^{j}(\mathbf{x}) \tilde{T}^\alpha_A \quad \mathbf{x} \notin A.
\end{cases}
\ee
We are now ready to compute the ratio of charged moments, along the same lines of the previous computation. Namely, given  a fermionic field $\Psi(\mathbf{x})$ with $U(1)$ charge $+1$, we consider the state
\be\label{eq:Exc_state_ferm}
| \boldsymbol{1}\ket \sim \Psi(\mathbf{p})| 0 \ket,
\ee
and its replicated version
\be
| \boldsymbol{1}\ket^{n} \sim \Psi^1(\mathbf{p})\dots \Psi^n(\mathbf{p})| 0 \ket.
\ee
Given two disconnected regions $A$ and $B$, we express the ratio of charged moments of the partial transpose also in this case as
\be
 \frac{{}^{n}\bra \boldsymbol{1} | T^\alpha_A \tilde{T}^\alpha_B| \boldsymbol{1} \ket^n}{{}^{n}\bra 0 | T^\alpha_A \tilde{T}^\alpha_B| 0 \ket^{n}},
\ee
with $n$ an even integer. We expand the expectation value of the twist operators as follows
\be
\begin{split}\label{eq:ManyTerms_ferm}
^n\bra 0 |(\Psi^\dagger)^n(-\mathbf{p})\dots(\Psi^\dagger)^1(-\mathbf{p})  (\Psi^2_A(\mathbf{p})+\Psi^n_B(\mathbf{p})e^{-2\pi i \alpha}+\Psi^1_{C}(\mathbf{p}))\dots \times\\
(-\Psi^1_{A}(\mathbf{p})e^{2\pi i \alpha} - \Psi^{n-1}_{B}(\mathbf{p})e^{2\pi i \alpha}+ \Psi^n_{C}(\mathbf{p}))T^\alpha_A \tilde{T}^\alpha_B|0\ket^n \simeq \\
^n\bra 0 |(\Psi^\dagger)^n(-\mathbf{p})\dots(\Psi^\dagger)^1(-\mathbf{p})  (\Psi^2_A(\mathbf{p})+\Psi^n_B(\mathbf{p})e^{-2\pi i \alpha}+\Psi^1_{C}(\mathbf{p}))\dots \times\\
(-\Psi^1_{A}(\mathbf{p})e^{2\pi i \alpha} - \Psi^{n-1}_{B}(\mathbf{p})+ \Psi^n_{C}(\mathbf{p}))|0\ket^{n} \times {}^{n}\bra 0 |T^\alpha_A \tilde{T}^\alpha_B|0\ket^n.
\end{split}
\ee
As in the bosonic case, many terms are generated after the sums are expanded and similar considerations as to which are leading and which are sub-leading for large volume can be applied. However, since the fermionic fields anticommute, we need to pay attention to the order of the fields. For example, the term $(A \ A \dots A)$ can be evaluated to
\be
\begin{split}
^n\bra 0 |(\Psi^\dagger)^n(-\mathbf{p})\dots(\Psi^\dagger)^1(-\mathbf{p})\Psi^2_A(\mathbf{p})\dots \Psi^1_A(\mathbf{p})|0\ket^n (-e^{2\pi i \alpha}) = \\
^n\bra 0 |(\Psi^\dagger)^n(-\mathbf{p})\dots(\Psi^\dagger)^1(-\mathbf{p})\Psi^1_A(\mathbf{p})\Psi^2_A(\mathbf{p})\dots \Psi^n_A(\mathbf{p})|0\ket^n e^{2\pi i \alpha} \sim e^{2\pi i \alpha} V^n_A,
\end{split}
\ee
where $\Psi^1_A(\mathbf{p})$ has been recast in the first position after crossing $n-1$ (odd) fermions, thus acquiring an additional phase $-1$. Similarly, it is easy to show that
\be
^n\bra 0 |(\Psi^\dagger)^n(-\mathbf{p})\dots(\Psi^\dagger)^1(-\mathbf{p})\Psi^n_B(\mathbf{p})\dots \Psi^{n-1}_B(\mathbf{p})|0\ket^n (-e^{-2\pi i \alpha}) \sim V^n_B e^{-2\pi i \alpha}.
\ee
So far, it should be clear that each term of the expansion is weighted with a proper phase, which arises from both the commutation relations (between twist operators and fermions) and the contractions, and this is the crucial difference with respect to the calculation presented for the boson. We can summarise the total contribution to the phase for a generic term 
\beq 
\bigl(\begin{smallmatrix}
 A_1 & \dots &  A_n\\
 j_1 & \dots & j_n \\
\end{smallmatrix}\bigr):= \,^n\bra 0 |(\Psi^\dagger)^n(-\mathbf{p})\dots(\Psi^\dagger)^1(-\mathbf{p})\Psi^{j_1}_{A_1}(\mathbf{p})\dots \Psi^{j_n}_{A_n}(\mathbf{p})|0\ket^n
\eeq 
as follows:
\begin{itemize}
    \item If $j_n =1$ and $A_n=A$ there is a $-e^{2\pi i\alpha}$ phase. Similarly, if $j_1 = 1$, and $A_1=B$, there is a contribution of $-e^{-2\pi i\alpha}$.
    \item In addition to the previous phase, an additional $-1$ is present for each $B$ which appears in the string $(A_{2} \ \dots \  A_n)$. This is due to the fermionic partial transposition over $B$.
    \item Once the contraction is performed, there is a sign coming from the order of the fields. Given $ (j_1  \ \dots \ j_n) = (\sigma(1) \dots \sigma(n))$, with $\sigma$ a generic permutation of the replica indices, one can show that the sign appearing after the contraction is $\text{sign}(\sigma)$.
\end{itemize}
Having suitably modified the definition of the twist operators for the fermion, the phase of each term appearing in \eqref{eq:ManyTerms_ferm} is the same as the one for the corresponding term in \eqref{eq:manyTerms}, leading to the same result for fermions and bosons. To see this, let us first analyse a term of type-I
\be
\bigl(\begin{smallmatrix}
 B & A_2 \dots &  A\\
 n & j_2 \dots & 1 \\
\end{smallmatrix}\bigr).
\ee
The phases coming from the last $A$ and the first $B$ cancel each other. Let $A_{i+1}=B$: then this must be preceded by $A_{i} =A$. The resulting replica indices at the corresponding positions are $j_{i+1} = i$ and $j_i = {i+1}$. In other words, there is an exchange of the replica indices $i$ and $i+1$, which changes the sign of the permutation $\sigma$ and it contributes as $-1$ after the contraction, but it is compensated by the $-1$ due to the presence of a $B$. Similar considerations apply straightforwardly to the strings of type-II. 
Putting everything together, the same formula \eqref{eq:ratio_ch_mom_abs} is obtained again, which is the final result. We emphasise that this derivation relies on the assumption that $n$ is even, something that was not necessary for a single bosonic excitation. 

\subsubsection{Single Fermionic Excitation via Replica Diagonalisation}\label{sec:Fermions_free}

We briefly note that the same result just derived for fermions can also be obtained via replica diagonalisation in the free case. The key observation is that, as show in Ref. \cite{ssr-17}, the fermionic partial transpose of a Gaussian state is still Gaussian. This allows to simplify the analysis of the replica theory, reducing it to a single-replica model in the presence of proper fluxes. For instance, given a Gaussian state and its RDM, and taking $n$ to be even, one can show the factorisation \cite{ssr-17}
\be\label{eq:product_ferm_Gauss}
\text{Tr}\l |\rho_{A\cup B}^{T_B}|^{n}e^{2\pi i \alpha (Q_A - Q_B) }\r = \prod^{\frac{n-1}{2}}_{p=-\frac{n-1}{2}} \text{Tr}\l \rho_{A\cup B}e^{\frac{2\pi i (\alpha+p)}{n} (Q_A - Q_B) + i\pi Q_B } \r.
\ee
Each term appearing inside the product is nothing but a single copy charged partition function with given fluxes along $A$ and $B$, that is the key quantity we aim to evaluate in this subsection. 

we can also identify each term inside the product (\ref{eq:product_ferm_Gauss}) as a ratio of correlators of twist operators, as done in \ref{eq:ratio_ch_moments}.
The only difference is the flux $\alpha$ in $T_A^\alpha$, which needs to be shifted in accordance with the form of the trace inside the product. 
We then have that 
\beq 
\mathcal{R}_1^n(r_A,r_B,r;\alpha)= \prod^{\frac{n-1}{2}}_{p=-\frac{n-1}{2}} \frac{\text{Tr}\l \rho_{A\cup B}e^{\frac{2\pi i (\alpha+p)}{n} (Q_A - Q_B) + i\pi Q_B } \r}{\text{Tr}\l \rho_{A\cup B,0}e^{\frac{2\pi i (\alpha+p)}{n} (Q_A - Q_B) + i\pi Q_B } \r}=\prod^{\frac{n-1}{2}}_{p=-\frac{n-1}{2}} \frac{\bra \boldsymbol{1} | T^{\alpha+p}_A T^{-\alpha-p+\frac{1}{2}}_B| \boldsymbol{1} \ket}{\bra 0 | T^{\alpha+p}_A T^{-\alpha-p+\frac{1}{2}}_B| 0 \ket}\,,
\label{prodn}
\eeq
where the twist operators are defined by Eq. \eqref{eq:ferm_twist} with $n=1$. A similar calculation as in the previous sections gives 
\be
\begin{split}
\bra \boldsymbol{1} | T^\alpha_A T^{-\alpha+\frac{1}{2}}_B| \boldsymbol{1} \ket = \frac{\bra 0 |\Psi^\dagger(\mathbf{p}) T^{\alpha}_A T^{-\alpha+\frac{1}{2}}_B\Psi(\mathbf{p})| 0 \ket}{\bra 0 |\Psi^\dagger(\mathbf{p}) \Psi(\mathbf{p})| 0 \ket} =\\ \frac{\bra 0 |\Psi^\dagger(\mathbf{p}) (\Psi_A(\mathbf{p})e^{2\pi i\alpha} - \Psi_B(\mathbf{p})e^{-2\pi i\alpha}+\Psi_{C}(\mathbf{p})) T^{\alpha}_A T^{-\alpha+\frac{1}{2}}_B| 0 \ket}{\bra 0 |\Psi^\dagger(\mathbf{p}) \Psi(\mathbf{p})| 0 \ket} \simeq \\
\bra 0 |T^{\alpha}_A T^{-\alpha+\frac{1}{2}}_B| 0 \ket(r_A e^{2\pi i \alpha} - r_B e^{- 2\pi i\alpha} +r).
\end{split}
\ee
so that 
\be
\prod^{\frac{n-1}{2}}_{p=-\frac{n-1}{2}} \frac{\bra \boldsymbol{1} | T^{\alpha+p}_A T^{-\alpha-p+\frac{1}{2}}_B| \boldsymbol{1} \ket}{\bra 0 | T^{\alpha+p}_A T^{-\alpha-p+\frac{1}{2}}_B| 0 \ket} = \prod^{\frac{n-1}{2}}_{p=-\frac{n-1}{2}} (r_A e^{\frac{2\pi i (\alpha+p)}{n}} - r_B  e^{-\frac{2\pi i (\alpha+p)}{n}}+r)\,.
\ee
This product can be shown yet again to be equal to \eqref{eq:ratio_ch_mom_abs}, though the proof requires some mathematical identities that we present in Appendix \eqref{sec:Math_Id}. 

\subsection{Bosonic State with Multiple Distinct Excitations}

Consider now a $k$-particle bosonic state where all particles have the same (unitary) charge and momentum $\mathbf{p}$. In order to ensure the presence of $U(1)$ symmetry we consider the complex free boson, described by a field $\mathcal{O}$ which satisfies Wick's theorem in the vacuum state. We thus describe the excited state as
\be
| \boldsymbol{k}\ket \sim \l\mathcal{O}(\mathbf{p})\r^k| 0 \ket.
\ee
Before entering the core of the computation in the replica model it is convenient, for computational reasons, to slightly modify the definition \eqref{eq:bos_twist} of the twist operators as follows
\be
T^\alpha_A \mathcal{O}^j(\mathbf{x}) = \begin{cases} e^{\frac{2\pi i \alpha}{n}} \mathcal{O}^{j+1}(\mathbf{x}) T^\alpha_A \quad \mathbf{x} \in A, \\
\qquad  \quad \,\mathcal{O}^{j}(\mathbf{x}) T^\alpha_A \quad \mathbf{x} \notin A.
\end{cases}
\label{exnew}
\ee
This amounts to distributing the total flux $e^{2\pi i \alpha}$ among all copies, rather than inserting it between the $n$-th and the first replica only. While this operator is different from what we used before, one can show that the final result obtained as expectation value is not modified, as a consequence of replica and $U(1)$ symmetry. An analogous fractionalisation will be considered for $\tilde{T}^\alpha_A$. We can now evaluate
\be\label{eq:kPArt_ratio}
\begin{split}
{}^{n}\bra \boldsymbol{k} | T^\alpha_A \tilde{T}^\alpha_B| \boldsymbol{k} \ket^n = \frac{^n\bra 0 |\l(\mathcal{O}^\dagger)^n(-\mathbf{p})\r^k\dots\l(\mathcal{O}^\dagger)^1(-\mathbf{p})\r^k T^\alpha_A \tilde{T}^\alpha_B \l\mathcal{O}^1(\mathbf{p})\r^k\dots \l\mathcal{O}^n(\mathbf{p})\r^k|0\ket^n}{^n\bra 0 |\l(\mathcal{O}^\dagger)^n(-\mathbf{p})\r^k\dots\l(\mathcal{O}^\dagger)^1(-\mathbf{p})\r^k \l\mathcal{O}^1(\mathbf{p})\r^k\dots \l\mathcal{O}^n(\mathbf{p})\r^k|0\ket^n}.
\end{split}
\ee
The denominator, required to ensure the normalisation of the state, can be computed using Wick's theorem, which gives the expectation value as a sum over the possible contractions:
\be
\begin{split}
^n\bra 0 |\l(\mathcal{O}^\dagger)^n(-\mathbf{p})\r^k\dots\l(\mathcal{O}^\dagger)^1(-\mathbf{p})\r^k \l\mathcal{O}^1(\mathbf{p})\r^k\dots \l\mathcal{O}^n(\mathbf{p})\r^k|0\ket^n  =\\
[\bra 0 |\l(\mathcal{O}^\dagger)(-\mathbf{p})\r^k \l\mathcal{O}(\mathbf{p})\r^k|0\ket\big]^n\sim
V^{nk}(k!)^n.
\end{split}
\ee

The numerator in \eqref{eq:kPArt_ratio} can be manipulated similarly to the previous sections, now using the exchange relation (\ref{exnew}) :
\be
\begin{split}
^n\bra 0 |\l(\mathcal{O}^\dagger)^n(-\mathbf{p})\r^k\dots\l(\mathcal{O}^\dagger)^1(-\mathbf{p})\r^k  T^\alpha_A \tilde{T}^\alpha_B\l\mathcal{O}^1(\mathbf{p})\r^k\dots \l\mathcal{O}^n(\mathbf{p})\r^k|0\ket^n = \\
^n\bra 0 |\l(\mathcal{O}^\dagger)^n(-\mathbf{p})\r^k\dots\l(\mathcal{O}^\dagger)^1(-\mathbf{p})\r^k  \l\mathcal{O}^2_A(\mathbf{p})
e^{\frac{2\pi i \alpha}{n}} + \mathcal{O}^n_B(\mathbf{p})e^{-\frac{2\pi i\alpha}{n}} + \mathcal{O}^1_{C}(\mathbf{p})\r^k\dots \times\\
\l \mathcal{O}^1_A(\mathbf{p})e^{\frac{2\pi i \alpha}{n}} + \mathcal{O}^{n-1}_B(\mathbf{p})e^{-\frac{2\pi i\alpha}{n}} + \mathcal{O}^n_{C}(\mathbf{p})
\r^kT^\alpha_A \tilde{T}^\alpha_B|0\ket^n \simeq\\
^n\bra 0 |\l(\mathcal{O}^\dagger)^n(-\mathbf{p})\r^k\dots\l(\mathcal{O}^\dagger)^1(-\mathbf{p})\r^k  \l\mathcal{O}^2_A(\mathbf{p})e^{\frac{2\pi \alpha}{n}} + \mathcal{O}^n_B(\mathbf{p})e^{-\frac{2\pi i \alpha}{n}} + \mathcal{O}^1_{C}(\mathbf{p})\r^k\dots \times\\
\l \mathcal{O}^1_A(\mathbf{p})e^{\frac{2\pi i\alpha}{n}} + \mathcal{O}^{n-1}_B(\mathbf{p})e^{-\frac{2\pi i\alpha}{n}} + \mathcal{O}^n_{C}(\mathbf{p})
\r^k |0\ket^n \times  {}^n\bra 0 |T^\alpha_A \tilde{T}^\alpha_B|0\ket^n.
\end{split} 
\ee
The generic evaluation of the previous expression is combinatorially-speaking rather involved, however many crucial features are already apparent. Namely, it is clear that $3^{nk}$ terms are generated simply by expanding the product over sums. Each resulting term can then be evaluated via Wick's theorem and will give rise to many contractions, as there are many possible ways to contract $\mathcal{O}^j_{A_i}(\mathbf{p})$ with $(\mathcal{O}^\dagger)^j(-\mathbf{p})$. Any of these contractions can be recovered via the permutation of the operators $(\mathcal{O}^\dagger)^j(-\mathbf{p})$ living in the same replica, giving a factor $(k!)^n$ which precisely cancels the normalisation of the state (the denominator in Eq. \eqref{eq:kPArt_ratio}). Moreover, whenever the restriction of $\mathcal{O}^1_{A}(\mathbf{p})$ over the region $A$ appears, a factor $V_A e^{\frac{2\pi i \alpha}{n}}$ is present after the Wick contraction; similarly, a factor $V_B e^{-\frac{2\pi i \alpha}{n}}$ appears with every $B$ and a factor $V_{C}$ for every $C$. Putting everything together, we can infer the general structure
\be\label{eq:manyTerms_Bos}
\begin{split}
^n\bra 0 |\l(\mathcal{O}^\dagger)^n(-\mathbf{p})\r^k\dots\l(\mathcal{O}^\dagger)^1(-\mathbf{p})\r^k  \l\mathcal{O}^2_A(\mathbf{p})e^{2\pi i \alpha/n} + \mathcal{O}^n_B(\mathbf{p})e^{-2\pi i \alpha/n} + \mathcal{O}^1_{C}(\mathbf{p})\r^k\dots \times\\
\l \mathcal{O}^1_A(\mathbf{p})e^{2\pi i \alpha/n} + \mathcal{O}^{n-1}_B(\mathbf{p})e^{-2\pi i \alpha/n} + \mathcal{O}^n_{C}(\mathbf{p})
\r^k |0\ket^n =\\
(k!)^n \sum_{k_A,k_B}C_{k_A,k_B}V^{k_A}_AV^{k_B}_B(V_{C})^{nk-k_A-k_B}e^{\frac{2\pi i \alpha (k_A-k_B)}{n}},
\end{split}
\ee
so that the expectation value is a homogeneous polynomial of degree $nk$ in $V_A,V_{B},V_{C}$, and $C_{k_A,k_B}$ is a natural number of combinatorial nature. Using the expression Eq. \eqref{eq:ratio_ch_moments}, we get the ratio of charged moments as
\be
\mathcal{R}_k^n(r_A,r_B,r;\alpha) = \frac{{}^{n}\bra \boldsymbol{1} | T^\alpha_A \tilde{T}^\alpha_B| \boldsymbol{1} \ket^n}{{}^{n}\bra 0 | T^\alpha_A \tilde{T}^\alpha_B| 0 \ket^n} =
\sum_{k_A,k_B}C_{k_A,k_B}{r_A}^{k_A}{r_B}^{k_B}r^{nk-k_A-k_B}e^{\frac{2\pi i \alpha (k_A-k_B)}{n}}\,,
\ee
valid, as usual, in the infinite volume limit. 
The closed formula for the combinatorial coefficient $C_{k_A,k_B}$ at any $k$ and $n$ is difficult to obtain by this method, but has been obtained ealier for simpler qubit states. The coefficients $C_{k_A,k_B}$ are nothing but the coefficients $\mathcal{A}_{p,q}$ given in (\ref{apq}).

\subsection{An Example: $k=n=2$}
Let us consider the example $n=2$ and $k=2$ to get an idea of how the combinatorics of  \eqref{eq:manyTerms_Bos} works in this case. Define the symbol
\be
\bigl(\begin{smallmatrix}
 A_1 & \dots &  A_{nk}\\
 j_1 & \dots & j_{nk} \\
\end{smallmatrix}\bigr),
\ee
which represents to the insertion of a string $\mathcal{O}^{j_1}_{A_1}(\mathbf{p})\dots \mathcal{O}^{j_{nk}}_{A_{nk}}(\mathbf{p})$ inside the correlation function. Among the strings which are generated, we only keep those for which any replica index $j$ appears exactly $k$ times among $(j_1 \dots j_{nk})$, as all others will vanish after Wick contractions. For $n=k=2$ the length of the strings above is $nk=4$. For any given string, there are others that can be obtained via the following permutations of the $A_i$ indices:
\be\label{eq:symm_transf}
\begin{split}
&(A_1 \ A_2 \ A_3 \ A_4)\rightarrow (A_2 \ A_1 \ A_3 \ A_4), \quad (A_1 \ A_2 \ A_3 \ A_4)\rightarrow (A_1 \ A_2 \ A_4 \ A_3),\\
&(A_1 \ A_2 \ A_3 \ A_4)\rightarrow (A_3 \ A_4 \ A_1 \ A_2)
\end{split}
\ee
and these contribute equally. This allows us to slightly simplify the combinatorial counting, and we only list the strings up to the transformations generated by Eq. \eqref{eq:symm_transf}, taking care of the degeneracy for each representative distinct string. These are
\begin{itemize}
    \item $(A \ A \ A \ A)$ : it yields $(r_A e^{\frac{2\pi i \alpha}{2}})^4$.
    \item $(B \ B \ B \ B)$ : it yields $(r_B e^{-\frac{2\pi i \alpha}{2}})^4$.
    \item $(A \ A \ B \ B)$ and $(A \ B \ A \ B)$: they yield $6r^2_Ar^2_B$.
    \item $(B \ A \ A \ A)$: it yields $4(r_A e^{\frac{2\pi i \alpha}{2}})^3 (r_Be^{-\frac{2\pi i \alpha}{2}})$.
    \item $(A \ B \ B \ B)$: it yields $4(r_B e^{-2\pi i \alpha/2})^3 (r_Ae^{2\pi i \alpha/2})$.
    \item $(A \ C \ B \ C)$: it yields $8 r_A r_B r^2$.
    \item $(C \ C \ C \ C)$: it yields $r^4$.
    \item $(A \ C \ A \ C)$: it yields $4r^2(r_A e^{\frac{2\pi i \alpha}{2}})^2$.
    \item $(B \ C \ B \ C)$: it yields $4r^2(r_B e^{-\frac{2\pi i \alpha}{2}})^2$.
\end{itemize}
Putting all these pieces together, we obtain
\be
\begin{split}
\mathcal{R}_2^2(r_A,r_B,r;\alpha)= r^4_A e^{4\pi i\alpha}+r^4_B e^{-4\pi i\alpha} + 6 r^2_A r_B^2 + 4r^3_Ar_B e^{2\pi i \alpha} + 4r^3_Br_A e^{-2\pi i \alpha} +\\
8r_A r_B r^2 + r^4 + 4r^2_A r^2 e^{2\pi i \alpha} + 4r^2_B r^2 e^{-2\pi i \alpha}.
\end{split}
\ee
This result is consistent with the one obtained from a direct evaluation of the right hand side of \eqref{manyk} for $n=2$, $k=2$.

\section{Numerics}\label{sec:Numerics}

In this section, we present numerical results for a 1D lattice Fermi gas. In particular, we consider the ground-state at half-filling, which is a Fermi sea and has critical features. This Fermi sea is then excited through the insertion of an additional particle above the Fermi energy at large momentum. We aim to compute the ratio of charged moments for a state of a single excitation and show the validity of result \eqref{eq:ratio_ch_mom_abs} numerically. Agreement with the latter confirms the claim made earlier in this paper and in previous works, namely that while the ground-state exhibits theory-dependent, highly non-trivial behavior (in this case, captured by a free fermion CFT \cite{murciano2021symmetry}), the contribution given by the excitation is universal.

\subsection{The Method}

Let us consider a Fermi chain of length $L$ described by the fermionic operators $\{f_j,f^\dagger_j\}_{j=1,\dots,L}$ satisfying the standard anticommutation relations
\be
\{f_j,f_{j'}\} = \{f_j^\dagger ,f^\dagger _{j'}\} = 0, \qquad \{f_j,f^\dagger_{j'}\} = \delta_{jj'}.
\ee
We choose a Gaussian state with a given number of particles and consider its correlation matrix, denoted by
\be
C(j,j') = \la f^\dagger_j f_{j'} \ra, \quad j,j'=1,\dots,L.
\ee
Let us further define the $L\times L$ covariance matrix
\be
\Gamma= 1-2C\,.
\ee
Given any two disconnected spacial subsystems $A$ and $B$, of length $\ell_A,\ell_B$ respectively, the restriction of $\Gamma$ over $A\cup B$ is a $(\ell_A +\ell_B)\times (\ell_A +\ell_B)$ matrix defined by
\be
\Gamma_{A\cup B}=  \begin{pmatrix} \Gamma_{AA} & \Gamma_{AB}  \\ \Gamma_{BA}  & \Gamma_{BB}\end{pmatrix},
\ee

Following Ref. \cite{eisler2015partial} one can show that, if $\rho_{A\cup B}$ is the RDM for this system, the fermionic partial transposition $\rho^{R_B}_{A\cup B}$ is Gaussian. Moreover, since in general $\rho^{R_B}_{A\cup B}$ is not hermitian, it is convenient to introduce a matrix $\rho^{\times}$ defined as
\be
 \rho^{\times} = \frac{(\rho^{R_B}_{A\cup B})(\rho^{R_B}_{A\cup B})^\dagger}{ \text{Tr}(\rho_{A\cup B}^2)},
\ee
which, by construction, has unit trace. If one interprets $\rho^{\times}$ as an unphysical mixed state of $A\cup B$, its associated covariance matrix is (See \cite{eisler2015partial})
\be
\Gamma_{A\cup B}^\times\equiv \frac{2}{1+\Gamma^2_{A \cup B}} \begin{pmatrix} \Gamma_{AA} & 0 \\ 0 & -\Gamma_{BB}\end{pmatrix}.
\ee
One can then express the even charged moments of the partially transposed RDM as (See also Ref. \cite{murciano2021symmetry})
\be
\begin{split}
\log \text{Tr}\l |\rho^{R_B}_{A\cup B}|^{n} e^{2\pi i\alpha (Q_A-Q_B)} \r = \text{Tr} \log\l \l\frac{1-\Gamma^\times_{A\cup B}}{2}\r^{\frac{n}{2}}e^{2\pi i \alpha} + \l\frac{1+\Gamma^\times_{A\cup B}}{2}\r^{\frac{n}{2}} \r  \\ 
 + \frac{n}{2}\text{Tr} \log\l \l\frac{1+\Gamma_{A\cup B}}{2}\r^{2} + \l\frac{1-\Gamma_{A\cup B}}{2}\r^{2} \r.
 \end{split}
 \label{eq:num_neg_ch_moments}
\ee
We stress that Eq. \eqref{eq:num_neg_ch_moments} makes sense also if $n$ is not an even integer, and it naturally provides the analytic continuation over $n$ for the even charged moments.

\subsection{Lattice Fermi Gas}

For our numerics we take the Hamiltonian of a lattice free Fermi gas on a ring of length $L$
\be
    H = -\frac{1}{2}\sum_j f^\dagger_{j+1}f_j+f^\dagger_{j}f_{j+1}\,.
\ee
Its ground state is a Fermi sea, with Fermi momentum $k_F = \pi/2$, and its correlation matrix is
\be
C_0(j,j')\equiv \la f_j^\dagger f_{j'}\ra_{0} = \frac{\sin k_F(j-j')}{L \sin \frac{\pi(j-j')}{L}}\,.
\ee
We then consider the excited state obtained via the insertion of a particle at momentum
\be
k = k_F+\frac{\pi}{2}-\frac{\pi}{L}
\ee
above the Fermi sea, whose correlation matrix is
\be
C(j,j') = C_0(j,j') +\frac{1}{L}e^{-i(k_F +\frac{\pi}{2}-\frac{\pi}{L})(j-j')}\,.
\ee
While the specific choice of $k$ is irrelevant for our purpose, it is important to require $k-k_F$ finite in the thermodynamic limit\footnote{ We mention that in Ref. \cite{crc-20} the case $k-k_F \sim 1/L$, which is a low-lying state, was considered, and its symmetry resolved entanglement was computed.}

\begin{figure}[t]
\centering
	\begin{subfigure}{.5\textwidth}
	\includegraphics[width=\linewidth]{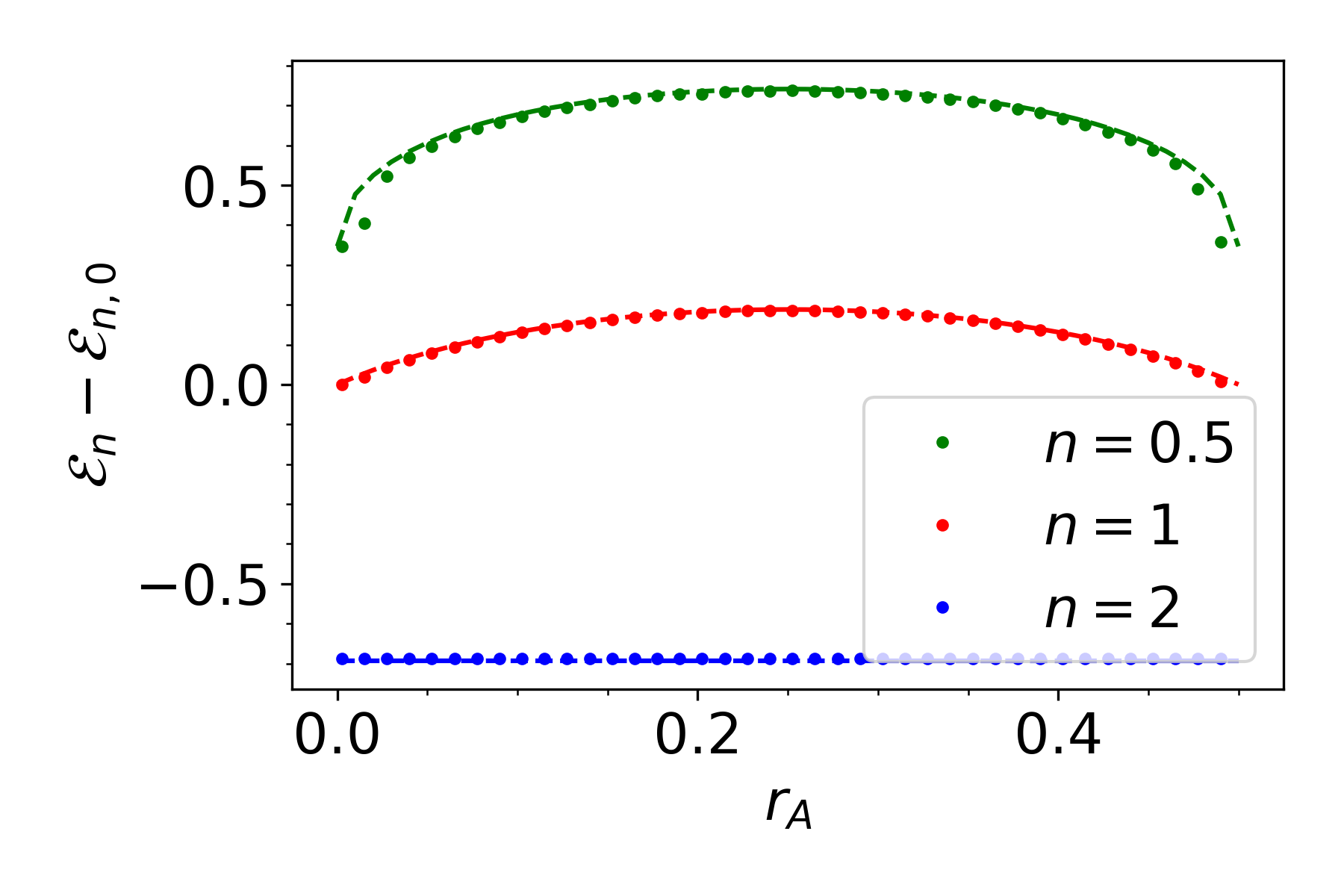}
	\end{subfigure}
    \caption{Difference of (uncharged, $\alpha = 0$) R\'enyi negativities for the one-particle state at $n = 0.5,1,2$. Note that although we derived our predictions for $n$ even via the replica approach, we can analytically continue them to any value of $n$ (and they are shown with dashed lines).}

    \label{fig:Neg}
\end{figure}

\begin{figure}[t]
	\begin{subfigure}{.5\textwidth}
	\includegraphics[width=\linewidth]{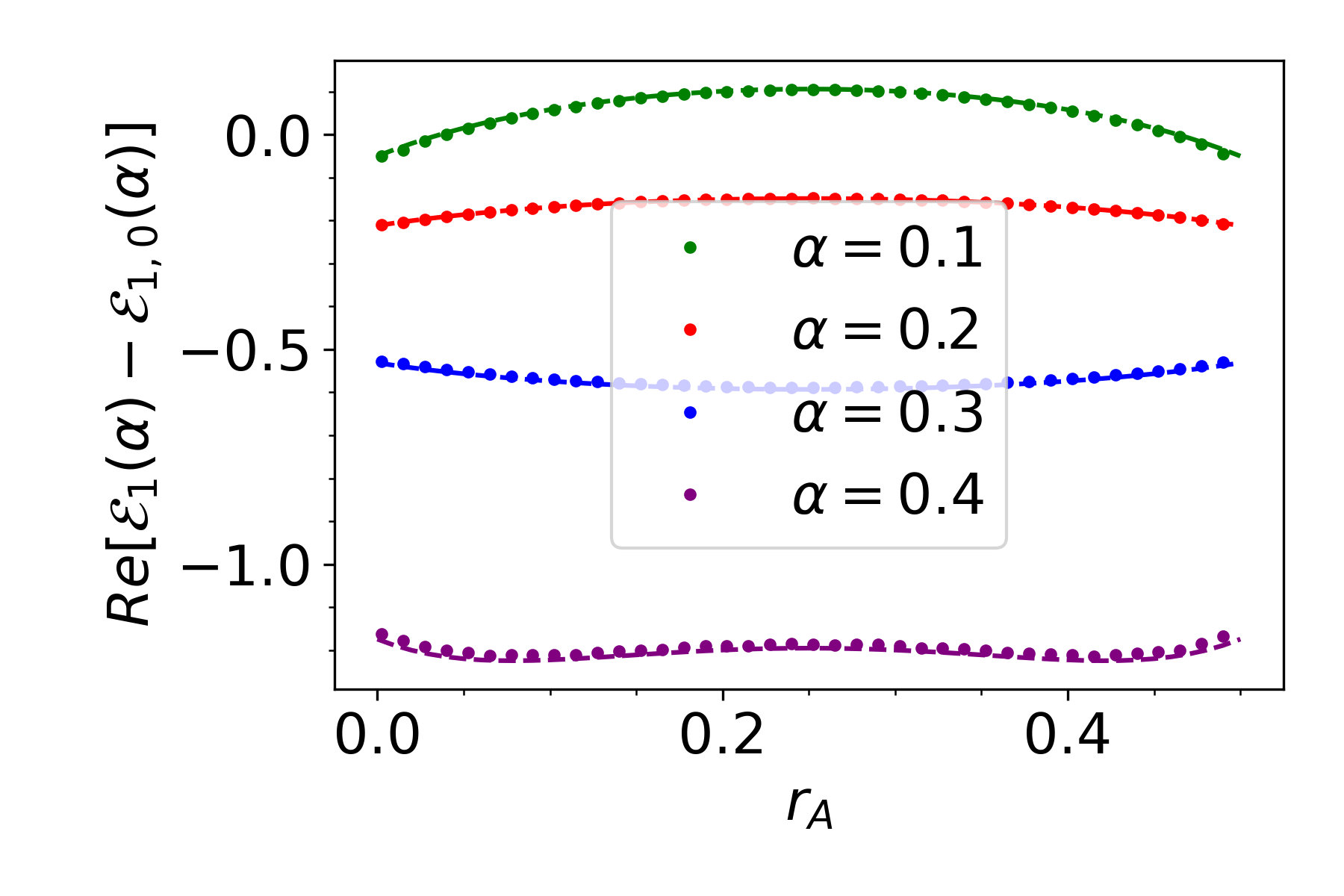}
	\end{subfigure}
	\begin{subfigure}{.5\textwidth}
	\includegraphics[width=\linewidth]{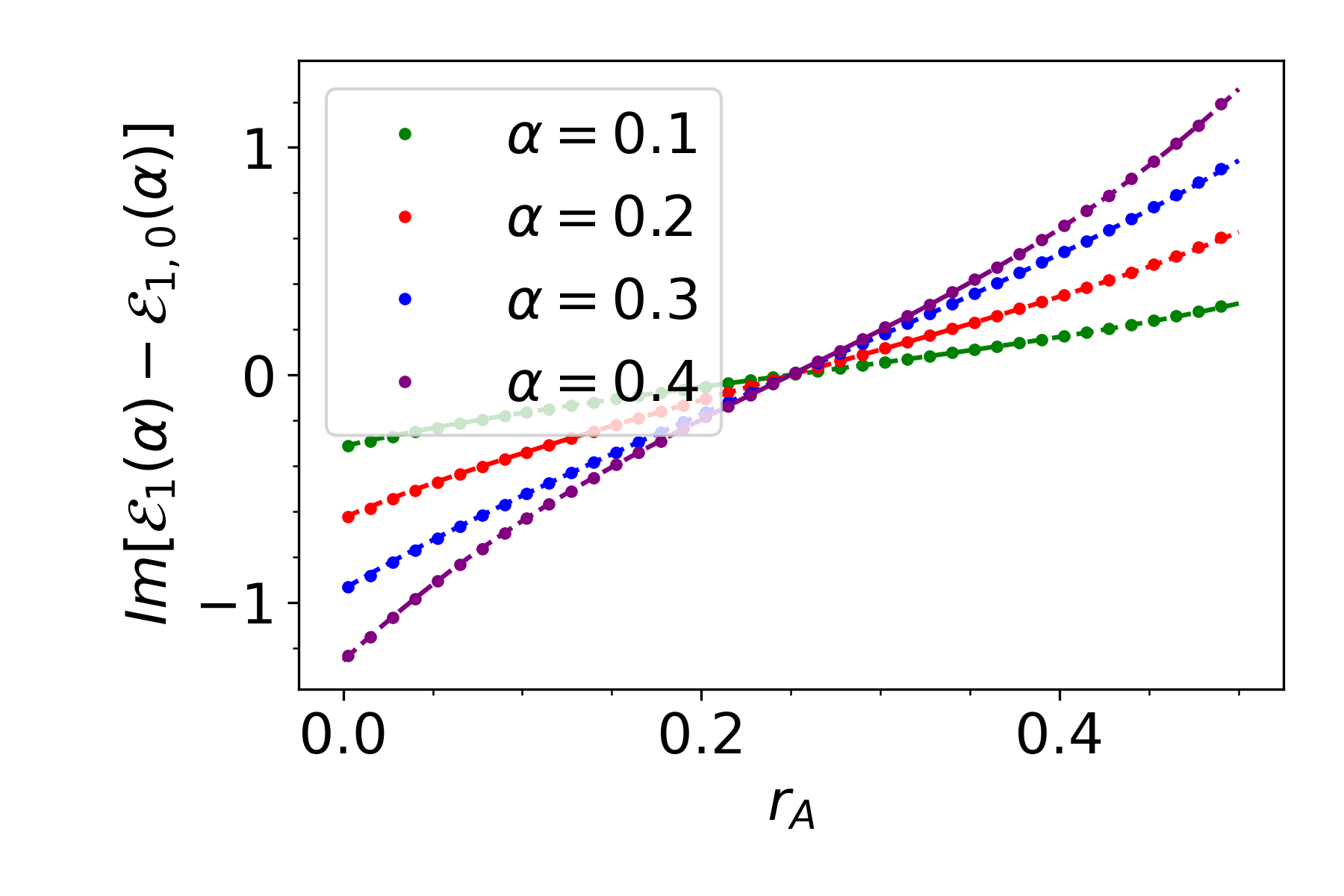}
	\end{subfigure}
	\begin{subfigure}{.5\textwidth}
	\includegraphics[width=\linewidth]{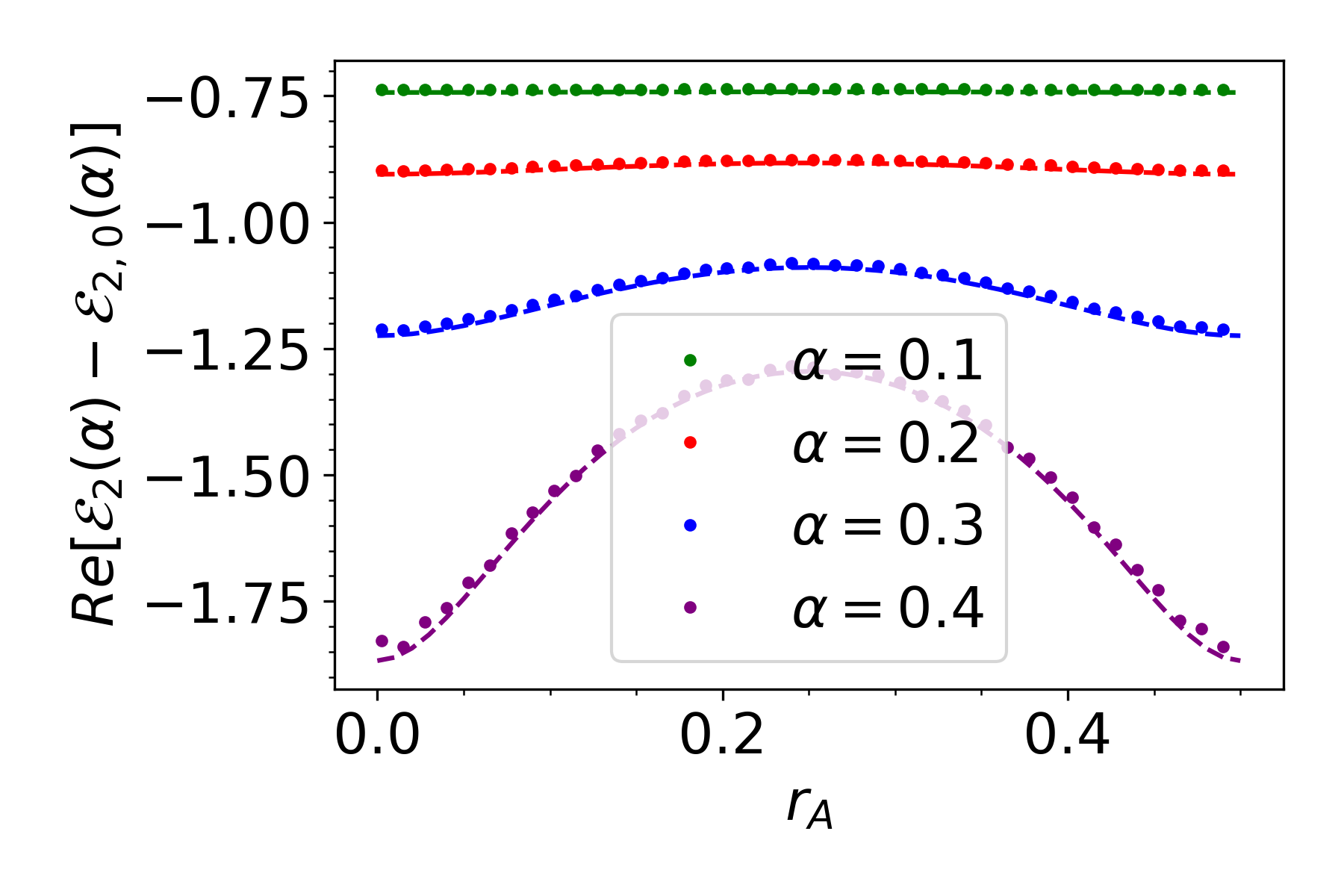}
	\end{subfigure}
	\begin{subfigure}{.5\textwidth}
	\includegraphics[width=\linewidth]{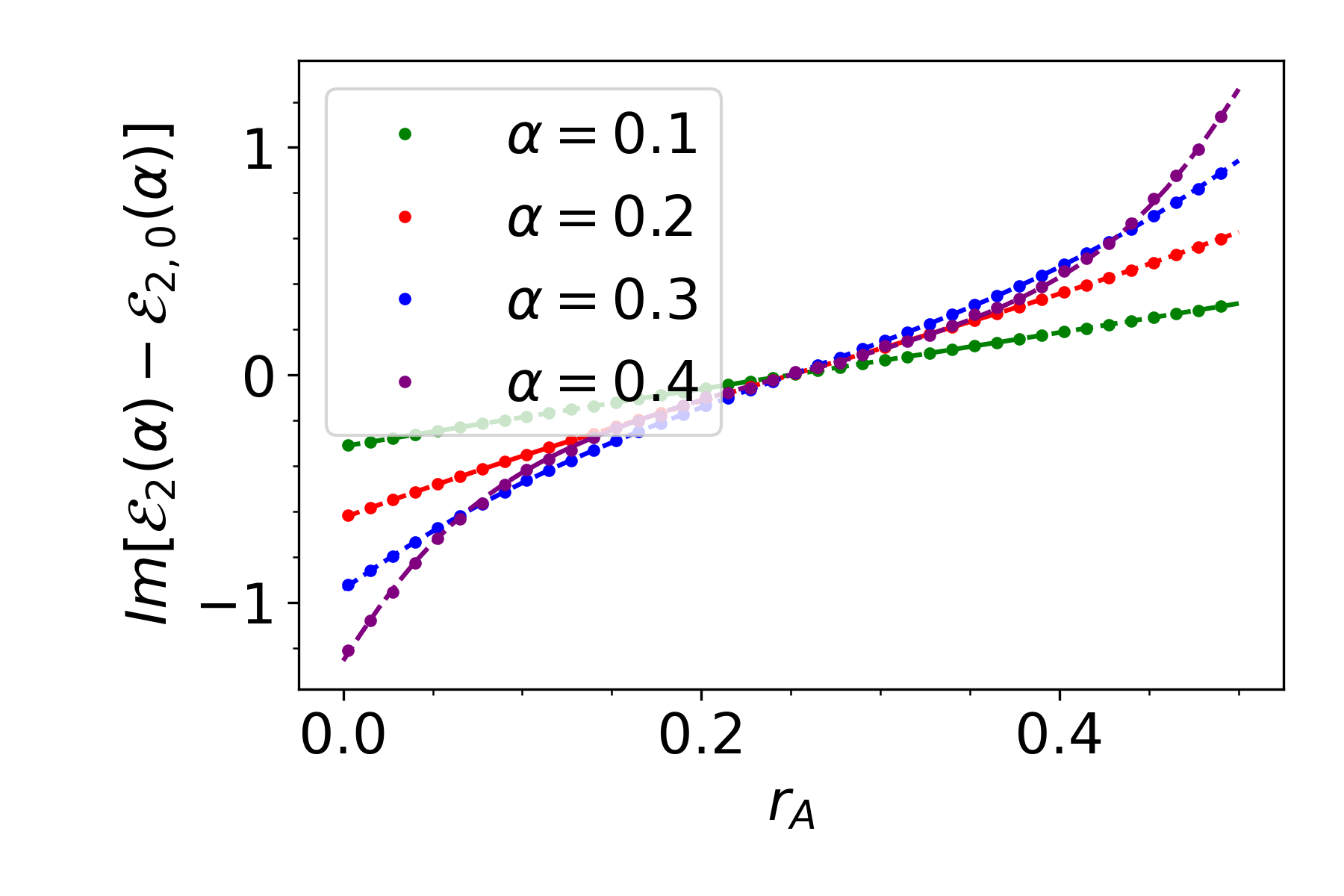}
	\end{subfigure}
    \caption{Difference of charged R\'enyi negativities for the one-particle state at flux $\alpha = 0.1,0.2,0.3,0.4$ and $n =1,2$ evaluated numerically (dots) versus the analytical predictions (dashed lines). The left/right panels show the real/imaginary part of $\mathcal{E}_{n}(\alpha) - \mathcal{E}_{n,0}(\alpha)$. The size of the chain is $L = 400$, and we plot the results as functions of $r_A \in (0,1/2)$. Note that although we have insisted that $n$ must be even, once a formula with $n$ even has been obtained, $n$ can be analytically continued to other values, hence our choices for the numerics.}
    \label{fig:ChNeg}
\end{figure}

We consider the following subsystems
\be
A = \{1,\dots,\ell_A\}, \quad B = \{\ell_A+1,\dots, \ell_A + \ell_B\},
\ee
and we fix the size of $A\cup B$ to be half the subsystem size:
\be
\frac{\ell_A+\ell_B}{L} = \frac{1}{2}\,.
\ee
We finally evaluate numerically the difference of charged R\'enyi negativities
\be
\mathcal{E}_{n}(\alpha) - \mathcal{E}_{n,0}(\alpha) \equiv \log \frac{\text{Tr}\l |\rho^{R_B}_{A\cup B}|^{n} e^{2\pi i \alpha (Q_A-Q_B)}\r}{\text{Tr}\l |\rho^{R_B}_{A\cup B,0}|^{n} e^{2\pi i \alpha (Q_A-Q_B)}\r},
\ee
for some values of the flux $\alpha$ as a function of $r_A =\ell_A/L$, and we compare it with the prediction \eqref{eq:ratio_ch_mom_abs} (with $r_B \equiv \ell_B/L = 1/2-r_A$). In Figs. \ref{fig:Neg} and \ref{fig:ChNeg} we show the results for a chain of length $L=400$ and given values of $n$ and $\alpha$, while varying the value of $r_A$ from 0 to $1/2$. We consider also non-even values of $n$, and we compare the numerics with the analytical continuation (over the even integers) of our predictions.

The general agreement is good: even if there are small discrepancies for small values of $\ell_A$ or $\ell_B$, corresponding to $r_A \simeq 0$ and $r_A \simeq 0.5$, these are expected to be finite-size effects which vanish in the large-volume limit.

\section{Conclusions and Outlook}
\label{conclusion}
This work is Part III of a series of papers, starting with \cite{partI, partII} where we have investigated the universal properties of symmetry resolved quantities in zero-density excited states. Zero-density here means that volume is taken to infinity, while the number of excitations above the ground state (which may be trivial, as for a qubit state, or highly non-trivial as in QFT) is kept fixed and finite. These papers in turn extend work on entanglement measures for zero-density excited states carried out in \cite{PRLexcited,castro2018entanglement,castro2019entanglement,graphpaper}. Other important contributions to this study are \cite{ZR1,ZR2,ZR3,ZR4,ZR5,ZR6,VM}

In line with the results of \cite{partI, partII} for the R\'enyi entropies, also here we expected and indeed found that the contribution of a finite number of excitations to the symmetry resolved (logarithmic) negativity is given by a simple formula, a polynomial on the variables $r_A, r_B$ and $r=1-r_A-r_B$, which represent the relative sizes of two subsystems $A$ and $B$ and their complement, respectively. For the symmetry resolved moments of the negativity, this polynomial will also depend on a parameter related to the internal symmetry of the theory, which in this paper we take to be $U(1)$. We called this parameter $\alpha$. The formulae that we obtained  generalise the results of \cite{castro2019entanglement,graphpaper} in a simple way and are consistent with numerical results. However, some of the methods that we have employed to obtain these results are quite new and have potential for further use. 

The method of twist operators, closely related to the derivation in \cite{graphpaper} for $d$-dimensional free bosonic theories, was introduced in \cite{partII} and the present paper provides a very non-trivial check of its validity. From this method alone, we can claim that our formulae should be valid in any dimensionality. Compared to a computation based on branch point twist fields for free QFT, as performed for the negativity in \cite{castro2019entanglement}, the use of twist operators, at least for free theories, captures the same universal result through a significantly simpler computation. 

A further application of twist operators and one of the most interesting and novel results of this paper is the fact that twist operators can be easily adapted to treat particles with both fermionic and bosonic statistics. In particular, it has been known for some time that the negativity of fermionic theories requires a slight redefinition of the operation of partial transposition \cite{eisler2016entanglement,ssr-17}. Here we find that, first, this redefinition is easy to implement in the context of twist operators, and, second, that once implemented it leads to a result which is the same as for bosons (even if the intermediate steps and starting point of the computation are different). This ties in well with the idea that the universal part of the entanglement associated with these types of excitation has a semiclassical interpretation (as recently explored in \cite{VM}), thus the statistics of excitations plays no role, even if it does play an important role for non-universal contributions, which are non-trivial when we consider excited states of QFT. 

\medskip

Looking ahead, there are many interesting questions to explore in relation to the role of different kinds of symmetries in the context of entanglement as well as the entanglement of excited states in different limits and for different types of particle statistics. 
For example, a notion of anyonic partial transposition was introduced in \cite{anyons} which we could easily adapt to some of the models considered here, such as qubit states.  There have also been recent studies of the symmetry resolved entanglement for non-invertible symmetries \cite{Lin} which could equally be extended to excited states. In connection to twist operators, any new measures of charged entanglement should be computable by a suitable redefinition of  the operators.

\medskip
Also related to this and previous work is the investigation of the crossover from low to high energy states in CFT, from the model-dependent predictions of \cite{ABS} to the sort of universal results obtained in \cite{PRLexcited,castro2018entanglement,castro2019entanglement,graphpaper,partI,partII} and here. The results of \cite{ABS} apply to low-lying excited states of CFT, whereas the universal formulae obtained for zero-density excited states apply for large momentum/energy. There must be a crossover between these two behaviours that one can understand from CFT arguments and it would be very interesting to do so. 
\medskip 

Finally, twist operators seem to be a promising  approach to computing entanglement measures in limiting cases where many details of the interaction can be neglected, i.e. semiclassical limit. A possible field where these ideas could be applied are out-of-equilibrium protocols \cite{pgb-22,Chen-22a}. It would be interesting to try to characterise entanglement growth for free/interacting theories in any dimension through this approach. For some protocols (i.e. global quench), we expect that the linear growth of entanglement may be captured by a semiclassical approximation of correlation functions, similar to what we have done here.
\medskip

We very much hope to tackle some of these problems in future work. 

\medskip

 \noindent {\bf Acknowledgements:} The authors are grateful to Cecilia De Fazio and Luc\'ia Sanz-Santamar\'ia for their involvement in Part I and Part II of this project and for many useful discussions that have contributed to the present work. Luca Capizzi thanks ERC for support under Consolidator grant number 771536 (NEMO) and the Department of Mathematics of City, University of London for hospitality during a two month visit (October-November 2022)  when most of this work was completed. Michele Mazzoni is grateful for funding under the EPSRC Mathematical Sciences Doctoral Training Partnership EP/W524104/1. Olalla A. Castro-Alvaredo thanks EPSRC for financial support under Small Grant EP/W007045/1.
 
\appendix

\section{Combinatorics}

\subsection{Non-Vanishing Strings}\label{sec:nonzero_strings}

Here, we count explicitly the terms in \eqref{eq:manyTerms} that give rise to non-vanishing contractions. In the following, we will make use of the notation introduced in \eqref{eq:string_notation}.
Let us consider a string containing the symbol $A$ at a certain position, say the first one, without loss of generality:
\be
\bigl(\begin{smallmatrix}
 A & A_2 &\dots &  A_n\\
 2 & j_2 & \dots & j_n \\
\end{smallmatrix}\bigr).
\ee
Then, the replica index $1$ should appear exactly once among $j_2,\dots,j_n$, as each of the replica indices $1,\dots,n$ has to be present if the corresponding term is non-vanishing. By inspecting \eqref{strings_J_i_values}, one realises that there are only two possible cases :
\begin{itemize}
\item $j_n = 1$, and then $A_n = A$,
\item $j_2 = 1$, which implies $A_2=B$.
\end{itemize}
If the first condition holds, one can apply the previous considerations to $A_n =A$ and deduce $A_{n-1} =A$; this argument can be iterated, and one concludes that $(A_1 \dots A_n ) = (A \ A \dots A)$. In contrast, if the second condition holds, then $\bigl(\begin{smallmatrix}
 A_1 & A_2 &\dots &  A_n\\
 j_1 & j_2 & \dots & j_n \\
\end{smallmatrix}\bigr) = \bigl(\begin{smallmatrix}
 A & B &\dots &  A_n\\
 1 & 2 & \dots & j_n \\
\end{smallmatrix}\bigr)$.

So far, as the choice of the first position was arbitrary (by cyclic permutation symmetry), we proved that whenever $A$ is present, either it is followed by $B$ or the whole string is $(A \ A \ \dots \  A)$. Using the same argument, one can prove that whenever $B$ is present, either it is preceded by $A$ or $(A_1 \ A_2 \ \dots \  A_n) = (B \ B \ \dots B)$.
\subsection{Combinatorial Counting of Strings}\label{sec:Comb_count}

In this Subsection we count the number of non-vanishing strings
\be
(A_1 \dots A_n)
\ee
containing $k$ pairs of consecutive $A$'s and $B$'s.
We first focus on the type-I strings
\be
(B \ A_2 \dots \dots A_{n-1} \ A).
\ee
The number of strings that satisfy the constraints derived above is given by all the possible ways one can insert sequences of $C$'s among any pair of $A$ and $B$. In other words, the generic string will look like
\be
(B \ C \ C\dots C \ A \ B \ C\ C\dots \dots C \ A),
\ee
where $k$ sequences of $C$'s of length $\{x_i\}_{i=1,\dots,k}$ are present, and $x_i\geq 0$ are integer numbers. As the length of the total string is $n$, the $\{x_i\}_{i=1,\dots,k}$ satisfy the following constraint
\be
x_1 +\dots +x_k = n-2k.
\ee
We now make use of a remarkable mathematical result, namely that the number of non-negative integer solutions of $x_1 +\dots +x_k = n$, that is the number of non-negative integer partitions of $n$ into $k$ parts is $\binom{n+k-1}{n}$ \cite{ntbook}. 
 As a consequence, the number of type-I strings satisfying the previous constraints is
\be
\binom{n-k-1}{k-1}.
\ee
Similarly, we consider now the type-II strings, having the following structure
\be
( C \ C\dots C \ A \ B \ C\ C\dots \dots C ).
\ee
In this case, there are $k+1$ sequences of consecutive $C$'s, as the difference with respect to the previous case is that there is now also a sequence that precedes the first pair of $A$ and $B$. Thus, we now have to count the number of non-negative integer solutions of
\be
x_1+\dots + x_{k+1} = n-2k,
\ee
which is
\be
\binom{n-k}{k}.
\ee
Summing up the contribution of both type of strings, we get precisely
\be
\binom{n-k-1}{k-1} + \binom{n-k}{k}
\ee
as the number of strings containing $k$ pairs of consecutive $A$'s and $B$'s.

As a last technical remark, we observe that there is at least one pair of consecutive $A$ and $B$ in the type-I strings, given by $A_n = A$ and $A_1 = B$. Then, if $k=0$ there are no strings satisfying the constraints, and this is compatible with the convention
\be
\binom{n-1}{-1} =0.
\ee

\section{Generalised Lucas Polynomials and a proof of Eq. \eqref{Lucas_main}}
\label{Lucas}
The generalised Lucas polynomials (in two variables) are defined via the recurrence relation \cite{bergum1974irreducibility,ricci1995generalized}:
\be
\label{Lucas_recurrence}
V_{n+2}(x,y) = x V_{n+1}(x,y) + y V_n(x,y)\,, \quad{n \in \mathbb{N}_0}
\ee
the first few polynomials are $V_0(x,y)=2$, $V_1(x,y)=x$, $V_2(x,y)= x^2+2y$. The proof of \eqref{Lucas_main} is based on the fact that the two sides of the equation are precisely two equivalent closed formulae for the $n$-th Lucas polynomial, with $x=r$, $y=r_A r_B$. In fact, we will now prove the following two statements:
\begin{enumerate}
    \item For all integers $n \ge 0$, one has a generalised Binet formula 
    \be
    \label{Lucas_Binet}
    V_{n}(x,y)= \alpha^n + \beta^n\,, \quad \alpha= \frac{x+\sqrt{x^2+4y}}{2}\,,\quad \beta = \frac{x-\sqrt{x^2+4y}}{2}\,. 
    \ee
    This is immediate to prove, as \eqref{Lucas_recurrence} holds by inspection for $n=0$, $n=1$ and furthermore $\alpha^2= x\alpha + y$, $\beta^2= x\beta + y$, which implies that $\alpha^{n+2}= x\alpha^{n+1}+ y\alpha^n$ and $\beta^{n+2}= x\beta^{n+1}+ y\beta^n$. This means that $\alpha^n+\beta^n$ satisfies the relation \eqref{Lucas_recurrence} for all $n \ge 0$.
    \item For all integers $n \ge 1$ another explicit formula for $V_n(x,y)$ is given by
    \be
    \label{Lucas_explicit}
    V_{n}(x,y)= \sum_{k=0}^{\rfloor n/2 \lfloor} \frac{n}{n-k}\binom{n-k}{k}x^{n-2k}y^{k}\,.
    \ee
    We prove this by showing again that the recurrence relation is satisfied. For $n=1$, $n=2$ it is immediate to see that this reproduces the correct polynomials. For $n \ge 3$, we can make use of the identity
    \be
    \label{Lucas_binomial_intermediate}
    \frac{n}{n-k}\binom{n-k}{k} = \frac{n-1}{n-k-1}\binom{n-k-1}{k}+\frac{n-2}{n-k-1}\binom{n-k-1}{k-1}
    \ee
    and we adopt the convention that $\binom{n}{k}=0$ if $k > n$ or $k < 0$. It is convenient to split the two cases $n=2m$ and $n=2m+1$, as the floor function yields different values. Let us consider the case $n=2m$, the other case being completely analogous. If $n=2m$, $\lfloor n/2 \rfloor = m$, $\lfloor (n-1)/2 \rfloor = \lfloor (n-2)/2 \rfloor = m-1$. From \eqref{Lucas_explicit} and \eqref{Lucas_binomial_intermediate} we have
    \begin{align}
    V_{n}(x,y)&= \sum_{k=0}^{m} \frac{2m}{2m-k}\binom{2m-k}{k}x^{2m-2k}y^{k} \nonumber \\
    &=x\,\sum_{k=0}^{m} \frac{2m-1}{2m-1-k}\binom{2m-1-k}{k}x^{2m-1-2k}y^{k} \nonumber \\ &+y\,\sum_{k=0}^{m} \frac{2m-2}{2m-1-k}\binom{2m-1-k}{k-1}x^{2m-2k}y^{k-1}
    \end{align}
    the first sum in the right-hand side vanishes if $k=m$, so that this term is $x V_{n-1}(x,y)$. The second sum on the other hand vanishes if $k=0$, so we can shift the summation variable and we see that this term reproduces $y V_{n-2}(x,y)$. Hence the recurrence relation is proved. 
\end{enumerate}
Equation \eqref{Lucas_main} follows from these two statements. This equation has (at least) two interesting implications. The first one comes from a direct expansion of the Binet formula using the binomial theorem:
\begin{align}
&\left(\frac{x-\sqrt{x^2+4y}}{2}\right)^n + \left(\frac{x+\sqrt{x^2+4y}}{2}\right)^n \nonumber \\
&=\frac{1}{2^n}\sum_{j=0}^n \left[(-1)^k  \binom{n}{j} x^{n-j}(x^2+4y)^{j/2}+\binom{n}{j} x^{n-j}(x^2+4y)^{j/2}\right]\nonumber \\
&=2^{1-n}\sum_{j=0}^{\lfloor n/2 \rfloor} \left[  \binom{n}{2j} x^{n-2j}(x^2+4y)^j\right]\nonumber \\
&=2^{1-n}\sum_{j=0}^{\lfloor n/2 \rfloor}  \binom{n}{2j} x^{n-2j}\sum_{k=0}^j \binom{j}{k} x^{2j-2k}2^{2k}y^k = 2^{1-n}\sum_{j=0}^{\lfloor n/2 \rfloor} \sum_{k=0}^j \binom{n}{2j}\binom{j}{k} x^{n-2k} 2^{2k}y^k\nonumber \\
&= 2^{1-n}\sum_{k=0}^{\lfloor n/2 \rfloor} x^{n-2k} 2^{2k}y^k \sum_{j=k}^{\lfloor n/2 \rfloor} \binom{n}{2j}\binom{j}{k} = \sum_{k=0}^{\lfloor n/2 \rfloor} \l 2^{1-n+2k}\sum_{j=k}^{\lfloor n/2 \rfloor}\binom{n}{2j}\binom{j}{k}\r x^{n-2k}y^k\,,
\end{align}
where in the last line we rearranged the sums over $j$ and $k$.
This quantity equals \eqref{Lucas_explicit}, which implies the non-trivial combinatorial identity:
\be
\label{byproduct_combinatorial}
\frac{n}{n-k}\binom{n-k}{k} = 2^{1-n+2k}\sum_{j=k}^{\lfloor n/2 \rfloor}\binom{n}{2j}\binom{j}{k}
\ee
As far as we know, this identity was only proved for $n$ odd in \cite{bergum1974irreducibility}.

The other interesting implication is obtained for $x=y=1$. In this case, the Lucas polynomials \eqref{Lucas_recurrence} reduce to the Lucas numbers:
\be
L_n = L_{n-1} + L_{n-2} \,, \quad n \ge 2
\ee
with $L_0 = 2$, $L_1=1$. The recurrence formula is the same defining the Fibonacci sequence, except for the different initial values. Equation \eqref{Lucas_Binet} with $x=y=1$ gives a closed formula for the Lucas numbers, and thus we have, for $n \ge 1$:
\be
\sum_{k=0}^{\rfloor n/2 \lfloor} \frac{n}{n-k}\binom{n-k}{k} = \l \frac{1+\sqrt{5}}{2} \r^n + \l \frac{1-\sqrt{5}}{2} \r^n
\ee
the quantity on the right-hand side is $\phi^n + (1-\phi)^n$, with $\phi$ the golden ratio, and it is always a positive integer. On the other hand, the quantity on the left hand side is the number of non-vanishing strings of type-I and type-II (out of a total of $3^n$ possible strings) obtained via the contraction methods discussed in Section \ref{sec:Twist_op}.

\section{Mathematical Identities}\label{sec:Math_Id}

Here we point out two useful identities which are employed to obtain the free fermion result in Subsection~\ref{sec:Fermions_free}. The first relation is
\be
\prod^{\frac{n-1}{2}}_{p=-\frac{n-1}{2}}(x+e^{\frac{2\pi i p}{n}}y) = x^n +y^n, \quad n \in \mathbb{N}
\ee
where the product is performed over $p$ integer/semi-integer when $n$ is odd/even. An immediate consequence of this identity is
\begin{align*}
&\prod^{\frac{n-1}{2}}_{p=-\frac{n-1}{2}}(xe^{\frac{2\pi i p}{n}}+ye^{-\frac{2\pi i p}{n}}+z) \\&= x^n \prod^{\frac{n-1}{2}}_{p=-\frac{n-1}{2}} \left[1 + \frac{z-\sqrt{z^2-4yx}}{2x}e^{-\frac{2\pi i p}{n}}\right] \left[1 + \frac{z+\sqrt{z^2-4yx}}{2x}e^{-\frac{2\pi i p}{n}}\right] \\&= x^n\left[1 + \left(\frac{z-\sqrt{z^2-4yx}}{2x}\right)^n\right]\left[1 + \left(\frac{z+\sqrt{z^2-4yx}}{2x}\right)^n\right] \\&= x^n + y^n + \left(\frac{z-\sqrt{z^2-4yx}}{2}\right)^n + \left(\frac{z+\sqrt{z^2-4yx}}{2}\right)^n.
\end{align*}
The latter equation can finally be employed to evaluate
\beqa
&& \prod^{\frac{n-1}{2}}_{p = -\frac{n-1}{2}}(r_A e^{\frac{2\pi i}{n}(\alpha+p)} - r_B e^{-\frac{2\pi i}{n}(\alpha+p)}+r)\nonumber \\
&& = e^{2\pi i \alpha} r^{n}_A + e^{-2\pi i \alpha}r^{n}_B +
\l \frac{r + \sqrt{r^2+4r_Ar_B}}{2}\r^{n} +\l \frac{-r + \sqrt{r^2+4r_Ar_B}}{2}\r^{n}\,,
\eeqa 
valid if $n$ an even integer \footnote{It is important here that $n$ is even, as $(-r_B)^{n} = r^{n}_B$.}.

\printbibliography
\end{document}